\documentclass[11pt]{article}

\usepackage[final]{acl}

\usepackage{times}
\usepackage{latexsym}

\usepackage[T1]{fontenc}
\usepackage[utf8]{inputenc}

\usepackage{microtype}

\usepackage{inconsolata}

\usepackage{graphicx}
\usepackage[table,xcdraw]{xcolor}

\usepackage{amsmath}
\usepackage{subcaption}

\title{When Vocal Tone and Literal Meaning Diverge: An Acoustic–Semantic Incongruity Study for Large Audio–Language Models}
\author{
\textbf{Yu-Wen Chen}$^{1,}$\thanks{These authors contributed equally. }
\quad
\textbf{William Ho}$^{1,}$\footnotemark[1] \quad
\textbf{Maxim Topaz}$^{2}$ \quad
\textbf{Zoran Kostic}$^{1}$ \quad
\textbf{Julia Hirschberg}$^{1}$ \\
$^{1}$The Fu Foundation School of Engineering and Applied Science, Columbia University, USA \\
$^{2}$School of Nursing, Columbia University, USA \\
\texttt{\{yu-wen.chen, william.ho\}@columbia.edu}
}

\begin{document}
\maketitle

\begin{abstract}
Affective cues across modalities may be incongruous (e.g., sarcasm or mocking praise), potentially leading to misinterpretation when relying on a single modality. Large Audio–Language Models (LALMs) have recently gained popularity and been applied to multimodal emotion recognition, but their ability to disentangle acoustic and semantic cues, especially in incongruent cases, remains underexplored. To address this gap, we introduce CREMA-ASIS, a dataset specifically created to investigate incongruence between acoustic emotion and semantic sentiment cues. It pairs acoustic emotion labels with semantic sentiment polarities. Using this dataset, we evaluate LALM biases within a multitask framework and conduct a layer-wise analysis to identify modality dominance across layers. Our findings reveal that LALMs struggle with semantic-acoustic incongruent cases, rarely predicting incongruity, and that LALMs are predominantly influenced by semantic information. However, supervised fine-tuning significantly improves LALM performance on our CREMA-ASIS test set while preserving transcription accuracy and joint emotion recognition. Results demonstrate potential for enhancing both acoustic and semantic understanding on out-of-domain data.
\end{abstract}

\section{Introduction}

Human affect is naturally conveyed through multiple modalities, including speech and text. Capturing emotional cues from both modalities provides deeper insight into the intent of the speaker~\citep{du2025unic}. Recent advances in Large Audio–Language Models (LALMs) have attracted increasing attention, enabling a unified framework supporting multiple audio-text tasks within a single model, including emotion recognition~\citep{chu2024qwen2, ding2025kimi, ghoshaudio, tang2024salmonn, deshmukh2023pengi}. However, most studies assign a single unified emotion label across modalities, thus overlooking both the heterogeneity and the complementary information provided by each. Moreover, researchers have observed that current LALMs may exhibit bias toward textual descriptions. For example,~\citet{wang2025audio} showed that LALMs tend to prioritize textual information in \emph{prompts} over acoustic evidence when the sources conflict. Beyond prompt-audio conflicts, mismatches can occur within the audio signal itself when acoustic and semantic content diverge. Such mismatches frequently occur in real-world communication, as in sarcasm, mocking praise, or concealed sadness. Dominance of one modality can bias interpretation, leading listeners to misread or overlook the speaker’s true intent.

Recent studies have begun exploring LALM bias in emotion recognition.~\citet{chen2025audio} investigated how different instructions (listening, reading, or both) affect LALM performance on emotion classification.~\citet{correa2025evaluating} evaluated LALMs using audio samples generated by a text-to-speech (TTS) system from GPT-generated sentences and speech emotion samples. Both studies observe a consistent lexical dominance. However, these studies mainly highlight LALMs’ bias toward semantic cues in audio, without further exploring potential strategies for mitigating this behavior. In addition, the datasets used in both works are limited, making them insufficient to guide models to disentangle acoustic and semantic cues. While previous studies focus on single-task settings, a more challenging yet practical approach is a multitask framework that generates separate predictions for each modality, enabling finer-grained analysis of speaker intent through comparison of modality-specific affective cues.

% not sure whether this revision is better. this is the paragraph that get complainted 

While prior work uses emotion categories or sentiment scores for both modalities, we consider pairing acoustic emotion categories (e.g., happy, angry) with semantic sentiment labels (positive, neutral, negative), which provides a more appropriate and informative setting. This choice was made to differentiate between a speaker’s emotion expressed through acoustic cues and the general consensus sentiment conveyed in the semantics. For example, “He had a car accident.” can express neutral (factual), sad (sympathetic), or even happy (malicious) emotions; however, for the lexical meaning, the general consensus sentiment of the event is negative. Acoustic signals are represented using emotion categories rather than sentiment to better capture the richness of human expressive states.

Our contributions are summarized as follows (1) We introduce CREMA-ASIS, an Acoustic–Semantic Incongruity Study dataset built upon CREMA-D~\citep{cao2014crema}. The dataset is created using a state-of-the-art (SOTA) text-to-speech (TTS) system, allowing controlled manipulation of acoustic and semantic labels and enabling systematic evaluation of how LALMs interpret multimodal affective cues. 
(2) We analyze the performance of several SOTA LALMs on CREMA-ASIS, revealing their strengths and limitations across various acoustic–semantic pairings. 
(3) Leveraging the scale of CREMA-ASIS, we conduct Supervised Fine-Tuning (SFT) experiments where models are trained to predict both acoustic emotion and semantic sentiment simultaneously. 
(4) We empirically demonstrate a pronounced acoustic–semantic gap in LALMs through layer-wise linear probing~\cite{alain2016understanding} and show that SFT on CREMA-ASIS improves the linear decodability of acoustic information in deeper layers while largely preserving semantic processing, narrowing the gap without sacrificing generalization. These findings provide diagnostic insight toward understanding how LALMs represent multimodal affective cues and highlight the potential of leveraging synthetic data to mitigate the acoustic–semantic gap\footnote{Code and data: \url{https://github.com/yuwchen/CREMA-ASIS}}.

%\begin{enumerate}

%\item We conduct layer-wise probing revealing the gap emerges from progressive acoustic information loss while semantic processing strengthens, and that fine-tuning enables acoustic retention throughout the model, providing mechanistic insight into improved performance.
%\end{enumerate}

\section{Method}

\begin{figure}[htbp!]
    \centering
\includegraphics[width=1.\columnwidth]{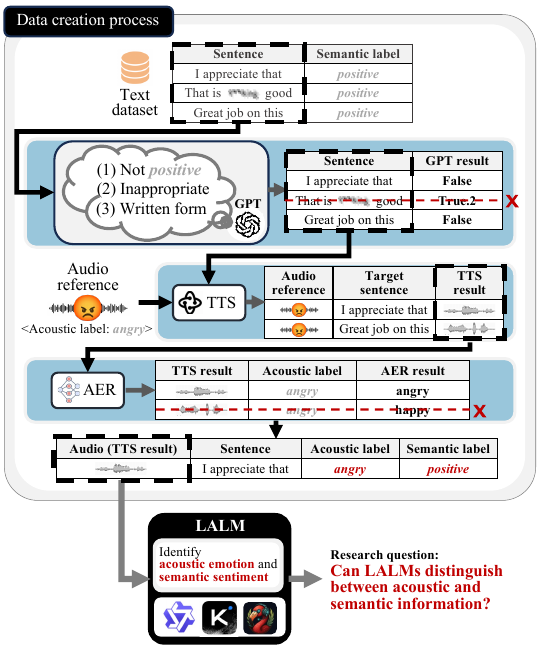}
    \caption{Study overview. The data creation process illustrates generating audio with an angry voice and positive semantic content.}
    \label{fig:method}
\end{figure}

We have designed CREMA-ASIS that explicitly targets affects which are inconsistent between acoustic and semantic information, specifically when the acoustic emotional valence contradicts the semantic sentiment expressed in the utterance. We used the dataset both to evaluate LALMs’ biases across acoustic and semantic modalities and to improve their ability to differentiate between the two via SFT. The models are instructed to perform acoustic emotion recognition and semantic sentiment prediction simultaneously (detailed in Appendix~\ref{sec:sft_prompt}). By explicitly pairing inconsistent acoustic and semantic signals, we create conditions where predicting acoustic emotion from lexical content alone yields systematically incorrect predictions. Successful learning therefore requires attending to and preserving acoustic features predictive of spoken emotion. Figure~\ref{fig:method} illustrates the study overview.

\subsection{CREMA-ASIS data creation}

Recent advances in TTS systems have enabled high-quality, natural-sounding speech generation. Among several open-source models supporting emotion expression and voice-cloning~\citep{du2024cosyvoice}, IndexTTS2~\citep{zhou2026indextts2} was selected based on manual listening tests. Its voice cloning capabilities allow it to replicate tone from the reference audio, enabling control of both acoustic emotion and semantic content in the generated speech by providing an emotion-specified reference sample.

\subsubsection{Audio Reference and Sentence Selection}
We focused on emotion categories that commonly exhibit acoustic emotion which is incongruous to semantic sentiment in real-world scenarios (Table~\ref{tab:scenario}). Specifically, a happy vocal tone paired with negative semantics often arises when speakers use a cheerful tone to complain, mask frustration, or make jokes. A disgusted or angry tone with positive semantic content may be delivered when praise or gratitude is expressed mockingly or as an ironic commendation. Lastly, speakers may convey positive messages using a sad tone to conceal personal distress and avoid causing concern for the listener. We also generated data under acoustic-neutral and semantic-neutral conditions for comparison. CREMA-ASIS covers five acoustic emotion categories: happy, neutral, angry, disgust, and sad, and three sentiment categories: negative, neutral, and positive.

\begin{table}[htbp!]
    \centering
\includegraphics[width=0.95\columnwidth]{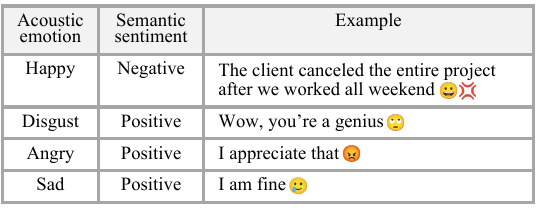}
    \caption{Acoustic–semantic incongruence scenarios.}
    \label{tab:scenario}
\end{table}

To generate controlled acoustic–semantic pairs, we required reference audio for each emotion category and sentences for each sentiment group. 

\paragraph{Audio Reference Selection:}
Speech emotion datasets include acted~\cite{burkhardt2005database} and spontaneous emotions~\cite{lotfian2017building, bujnowski2024samsemo}. While spontaneous emotions are more natural, we consider acted emotion datasets more appropriate for generating synthetic speech signals in this study. First, many spontaneous emotion datasets are derived from TV shows, where licensing restrictions often limit the distribution of derivative works. Second, most spontaneous datasets provide only overall labels, whereas generating audio references requires labels that specifically reflect the acoustic signal, independent of the semantic content. In contrast, acted datasets convey emotion solely through acoustic expression, keeping the semantic content neutral. In addition, acted emotions are often more exaggerated, making them easier for models to recognize. Failure to correctly classify speech synthesized from acted emotions indicates an even greater challenge in recognizing speech from spontaneous emotions. Therefore, we opted for an acted emotion dataset. Among these, CREMA-D~\footnote{Open Data Commons Attribution License (ODC-By) v1.0} was selected due to its relatively large number of unique speakers. Noticing some non-stationary noise in the dataset, we applied a speech enhancement model~\citep{chao2024investigation} to improve audio quality before using the samples as references in the TTS system.

\paragraph{Sentence Selection:} We did not use a Large-Language Model (LLM) to generate sentences because it often produces similar structures and repetitive words especially when more than a hundred of sentences are needed~\citep{holtzmancurious,yao2025understanding}. Instead, we used sentences from the GoEmotions\footnote{Apache License 2.0}~\citep{demszky2020goemotions} dataset, which are collected from Reddit and include human-annotated emotion categories. We focused on categories that express opinions, as acoustic–semantic incongruence often arises when people do not want to express their opinions directly to mitigate social tension, avoid offending others, or employ humor and irony~\cite{brown1987politeness}. For positive semantic sentiment, we selected the \emph{approval}, \emph{admiration}, and \emph{gratitude} categories. For negative semantic sentiment, we selected \emph{annoyance}, \emph{disapproval}, and \emph{disappointment}. Sentences in the \emph{neutral} category are also included to serve as control samples. We further used GPT-4o~\citep{hurst2024gpt} to filter the sentences based on the following criteria: (1) The sentence does not express the target sentiment. (2) The sentence contains content that is strongly racist, very offensive, or inappropriate for public use. (3) The sentence includes words or expressions that people typically would not say aloud (i.e., phrases that exist only in written form).

\subsubsection{Acoustic Filtering of Generated Speech}
We applied an acoustic-based emotion recognition (AER) model~\citep{yang2021superb} to identify and filter out samples whose conveyed emotion is unlikely to match the acoustic label. Although AER has limitations, it is effective in removing clear mismatches, e.g., when an angry target is rendered as sounding clearly happy, thereby improving the dataset’s overall emotional consistency.

\section{Experimental Settings}

\subsection{Synthetic Data Generation}

We use the GoEmotions dataset as the source of spoken content, randomly selecting a subset to limit experimental complexity and keep the size of synthetic data manageable. The resulting dataset consists of 2,255 negative, 1,944 neutral, and 2,267 positive sentences. These sentences are randomly split into training, validation, and test sets, and are randomly sampled during synthetic data generation. CREMA-D is used as the acoustic reference and consists of 7,442 speech clips produced by 91 actors. Each actor performs speech expressing basic emotional states, while the spoken content is restricted to 12 emotionally neutral sentences. We follow the original training, validation, and test splits provided with the dataset. The CREMA-D dataset is processed using SEMamba~\citep{chao2024investigation}. %~\footnote{\url{https://github.com/RoyChao19477/SEMamba/blob/main/ckpts/vd.pth}}. 
The AER is from Hugging Face~\footnote{firdhokk/speech-emotion-recognition-with-openai-whisper-large-v3}. Because AER is imperfect, we retained not only samples whose recognized emotion matches the target but also those with perceptually similar emotion categories (Appendix~\ref{sec:acoustic_filtering_settings}). After acoustic filtering, we apply \emph{whisper-base.en}~\citep{radford2023robust} to transcribe the generated speech and compare it with the specified spoken content. The resulting word error rate (WER) of 0.1 indicates that IndexTTS2 can reliably generate speech that closely matches the intended text in most cases. To further ensure evaluation reliability, we exclude samples with a WER greater than 0.5 in the test set.

\subsection{Supervised Fine-Tuning}\label{sec:experiment_setup_sft}
Three well-known LALMs were evaluated in this study: Qwen2-Audio~\citep{chu2024qwen2} (\emph{Qwen2-Audio-7B-Instruct)}, Kimi-Audio~\citep{ding2025kimi} (\emph{Kimi-Audio-7B-Instruct}), and Audio-Flamingo3~\citep{ghoshaudio}(\emph{audio-flamingo-3-hf}). In addition to CREMA-ASIS, we include the original acoustic-neutral pairs from the CREMA-D and MELD~\citep{poria2019meld} datasets during SFT. MELD, collected from TV series, exposes the model to more naturalistic emotional expressions, enhancing its generalizability. We use emotion and sentiment labels in MELD as acoustic emotion and semantic sentiment targets. Including these helps prevent the fine-tuned model from forgetting emotion categories it can initially recognize and from focusing solely on categories present in the synthetic data. SFT is performed using LoRA~\citep{hu2022lora, peft}, with parameter settings provided in Appendix~\ref{sec:SFT_setting}.

\subsection{Evaluation Metric}
We report acoustic accuracy ($Acc_{\text{acou}}$, emotion classification), semantic accuracy ($Acc_{\text{sem}}$, sentiment classification), and dual-condition accuracy ($Acc_{\text{dual}}$), which requires both predictions to be correct for a single instance. In addition to CREMA-ASIS and MELD test set, we evaluate generalizability on LISTEN\_full~\citep{chen2025audio}, where samples are annotated with either acoustic or semantic emotion categories. We use its test split\footnote{\url{https://huggingface.co/datasets/VibeCheck1/LISTEN\_full/blob/main/data/test-00000-of-00001.parquet}}. Since LISTEN\_full is collected from several open-source emotion datasets, we remove samples originating from CREMA-D and MELD to avoid domain overlap with our training data. Labels are manually mapped to the categories used in our dataset, and samples with ambiguous mappings are excluded from evaluation (details in Appendix~\ref{sec:listen_dataset_setup_Additional}). After filtering, we retain 1,123 samples for acoustic emotion analysis and 651 samples for semantic sentiment analysis. 

%For example, semantic surprise can convey either positive or negative sentiment and is therefore omitted. 

\subsection{Layer-wise Acoustic-semantic Probing}
To better understand how SFT affects acoustic vs. semantic information processing, we conduct layer-wise linear probing experiments~\citep{alain2016understanding} on the three LALMs from Section~\ref{sec:experiment_setup_sft} using CREMA-ASIS. We evaluate the discriminative power of hidden states across model depth for acoustic emotion ($Acc_{\text{acou}}$) and semantic sentiment ($Acc_{\text{sem}}$), and analyze the acoustic-semantic gap $G = |Acc_{\text{acou}} - Acc_{\text{sem}}|$. We compute changes between base and SFT models as $\Delta G = G^{\text{SFT}} - G^{\text{base}}$ (negative values indicate gap reduction), $\Delta Acc_{\text{acou}} = Acc_{\text{acou}}^{\text{SFT}} - Acc_{\text{acou}}^{\text{base}}$, and $\Delta Acc_{\text{sem}} = Acc_{\text{sem}}^{\text{SFT}} - Acc_{\text{sem}}^{\text{base}}$. %to attribute changes to specific modalities.

For each model, we extract representations from the audio encoder (Whisper), multi-modal projector, six LLM layers spanning early, middle and late depth ([1, 2, 9, 17, 25, 32] for Qwen2-Audio's 32 layers, [1, 2, 7, 14, 21, 28] for Kimi-Audio and Audio-Flamingo3's 28 layers), and Kimi-Audio's MIMO module. After mean pooling over the sequence dimension, we train linear classifiers to independently predict one of 5 acoustic or 3 semantic categories. Linear probes ensure results reflect information encoded in hidden states rather than classifier capacity. Training parameter settings are given in  Appendix~\ref{sec:experiment_setup_probing}.

\section{Results}

\subsection{CREMA-ASIS Dataset}
\subsubsection{Dataset Statistics and Composition}
Table~\ref{tab:dataset_statistic} presents CREMA-ASIS statistics: acoustic–semantic incongruous, congruous, and neutral-associated cases (where either or both the acoustic or semantic label is neutral). The incongruous pairs in which semantic sentiment opposes acoustic valence are challenging and common in real-world scenarios but underexplored in prior work. Congruous samples align acoustic and semantic cues, making them easier to recognize. Samples with exactly one neutral modality fall between incongruous and congruous, producing partially misaligned signals. Including congruous and neutral-associated samples helps prevent LALM from assuming acoustic and semantic cues are always inconsistent. Although the dataset distribution is slightly uneven due to post-processing, each category contains enough samples for effective SFT, so strict class balance was not enforced.
\begin{table}[htbp!]
\centering
\resizebox{0.9\columnwidth}{!}{%
\begin{tabular}{|cccccc|}
\hline
\multicolumn{2}{|c|}{}                                         & \multicolumn{1}{c|}{Train} & \multicolumn{1}{c|}{Val}  & \multicolumn{1}{c|}{Test} & Total \\ \hline
\multicolumn{2}{|c|}{Sample size}                              & \multicolumn{1}{c|}{65534} & \multicolumn{1}{c|}{6146} & \multicolumn{1}{c|}{5879} & 77559 \\ \hline
\multicolumn{2}{|c|}{\# of Actors}                             & \multicolumn{1}{c|}{63}    & \multicolumn{1}{c|}{14}   & \multicolumn{1}{c|}{14}   & 91    \\ \hline
\multicolumn{1}{|c|}{Acoustic} & \multicolumn{1}{c|}{Semantic} & \multicolumn{1}{c|}{}      & \multicolumn{1}{c|}{}     & \multicolumn{1}{c|}{}     &       \\ \hline
\multicolumn{6}{|c|}{\cellcolor[HTML]{EFEFEF}Acoustic-semantic Incongruous}                                                                                 \\ \hline
\multicolumn{1}{|c|}{disgust}  & \multicolumn{1}{c|}{positive} & \multicolumn{1}{c|}{4524}  & \multicolumn{1}{c|}{437}  & \multicolumn{1}{c|}{373}  & 5334  \\ \hline
\multicolumn{1}{|c|}{angry}    & \multicolumn{1}{c|}{positive} & \multicolumn{1}{c|}{6434}  & \multicolumn{1}{c|}{575}  & \multicolumn{1}{c|}{513}  & 7522  \\ \hline
\multicolumn{1}{|c|}{sad}      & \multicolumn{1}{c|}{positive} & \multicolumn{1}{c|}{3821}  & \multicolumn{1}{c|}{334}  & \multicolumn{1}{c|}{330}  & 4485  \\ \hline
\multicolumn{1}{|c|}{happy}    & \multicolumn{1}{c|}{negative} & \multicolumn{1}{c|}{3919}  & \multicolumn{1}{c|}{376}  & \multicolumn{1}{c|}{358}  & 4653  \\ \hline
\multicolumn{6}{|c|}{\cellcolor[HTML]{EFEFEF}Acoustic-semantic Congruous}                                                                                   \\ \hline
\multicolumn{1}{|c|}{disgust}  & \multicolumn{1}{c|}{negative} & \multicolumn{1}{c|}{4993}  & \multicolumn{1}{c|}{485}  & \multicolumn{1}{c|}{382}  & 5860  \\ \hline
\multicolumn{1}{|c|}{angry}    & \multicolumn{1}{c|}{negative} & \multicolumn{1}{c|}{7100}  & \multicolumn{1}{c|}{618}  & \multicolumn{1}{c|}{549}  & 8267  \\ \hline
\multicolumn{1}{|c|}{sad}      & \multicolumn{1}{c|}{negative} & \multicolumn{1}{c|}{3764}  & \multicolumn{1}{c|}{302}  & \multicolumn{1}{c|}{334}  & 4400  \\ \hline
\multicolumn{1}{|c|}{happy}    & \multicolumn{1}{c|}{positive} & \multicolumn{1}{c|}{3449}  & \multicolumn{1}{c|}{376}  & \multicolumn{1}{c|}{354}  & 4179  \\ \hline
\multicolumn{6}{|c|}{\cellcolor[HTML]{EFEFEF}Neutral-associated}                                                                                             \\ \hline
\multicolumn{1}{|c|}{disgust}  & \multicolumn{1}{c|}{neutral}  & \multicolumn{1}{c|}{2814}  & \multicolumn{1}{c|}{284}  & \multicolumn{1}{c|}{282}  & 3380  \\ \hline
\multicolumn{1}{|c|}{angry}    & \multicolumn{1}{c|}{neutral}  & \multicolumn{1}{c|}{3790}  & \multicolumn{1}{c|}{379}  & \multicolumn{1}{c|}{388}  & 4557  \\ \hline
\multicolumn{1}{|c|}{happy}    & \multicolumn{1}{c|}{neutral}  & \multicolumn{1}{c|}{2057}  & \multicolumn{1}{c|}{217}  & \multicolumn{1}{c|}{258}  & 2532  \\ \hline
\multicolumn{1}{|c|}{sad}      & \multicolumn{1}{c|}{neutral}  & \multicolumn{1}{c|}{4031}  & \multicolumn{1}{c|}{384}  & \multicolumn{1}{c|}{385}  & 4800  \\ \hline
\multicolumn{1}{|c|}{neutral}  & \multicolumn{1}{c|}{negative} & \multicolumn{1}{c|}{6108}  & \multicolumn{1}{c|}{534}  & \multicolumn{1}{c|}{530}  & 7172  \\ \hline
\multicolumn{1}{|c|}{neutral}  & \multicolumn{1}{c|}{neutral}  & \multicolumn{1}{c|}{3230}  & \multicolumn{1}{c|}{333}  & \multicolumn{1}{c|}{357}  & 3920  \\ \hline
\multicolumn{1}{|c|}{neutral}  & \multicolumn{1}{c|}{positive} & \multicolumn{1}{c|}{5500}  & \multicolumn{1}{c|}{512}  & \multicolumn{1}{c|}{486}  & 6498  \\ \hline
\end{tabular}%
}
  \caption{CREMA-ASIS Dataset statistics.}
  \label{tab:dataset_statistic}
\end{table}

\subsubsection{Subjective Human Validation of Acoustic Emotion Control}
\begin{figure}[hbtp!]
\centering
  \includegraphics[width=0.75\columnwidth]{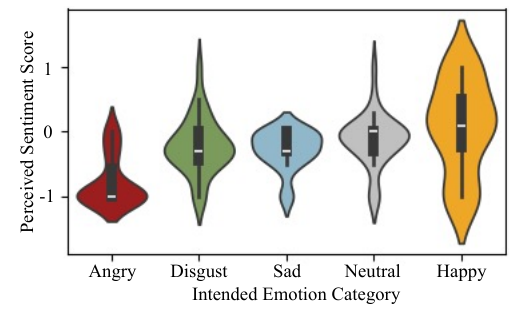}
  \caption{Results of subjective acoustic sentiment task.}
  \label{fig:listening_vio}
\end{figure}

\begin{figure}[hbtp!]
    \centering
  \includegraphics[width=0.98\columnwidth]{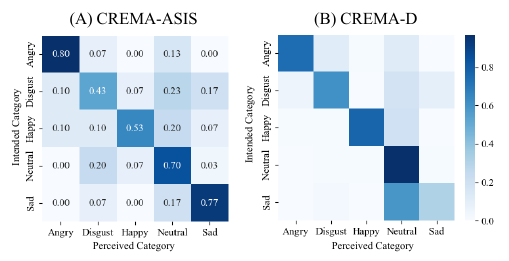}
  \caption{Confusion matrices of intended versus perceived emotions, showing similar patterns despite matrix B being based on audio-visual signals and matrix A on acoustic signals only. B is replotted from Figure 9 in~\citep{cao2014crema}; \emph{fear} is excluded to match CREMA-ASIS categories, and numerical values are omitted as they were computed with fear included.}
  \label{fig:listening_conf}
\end{figure}

We randomly sampled 150 instances (10 per category) from the test set for a subjective evaluation of the TTS system’s ability to reproduce the reference audio’s acoustics. Three evaluators performed two tasks: rating acoustic sentiment on a scale from –1 (negative) to 1 (positive) and assigning an acoustic emotion category. Figure~\ref{fig:listening_vio} shows violin plots of the intended emotions of the reference audio and their corresponding subjective sentiment scores, indicating that median perceived sentiment aligns with the intended acoustic valence (e.g., lowest for angry, highest for happy). Accuracy of acoustic categorical emotion recognition, computed using majority voting, is 64.6\% with inter-annotator agreement Krippendorff's $\alpha$ 0.54. Figure~\ref{fig:listening_conf} shows the intended-versus-perceived emotion confusion matrix of CREMA-ASIS and that reported from CREMA-D paper. The confusion patterns in our validation closely follow those observed in the original data. In particular, the neutral and disgust categories remain challenging: emotions with weak acoustic cues are often perceived as neutral and disgust is inherently difficult to recognize. In addition, when we listened to samples in which the perceived emotion diverged from the intended emotion of the reference audio, the synthetic and reference audio remained acoustically similar. Overall, these results suggest that the TTS system can reliably produce samples with controlled acoustic emotion and semantic sentiment.

\begin{figure*}[t]
\centering
  \includegraphics[width=0.95\textwidth]{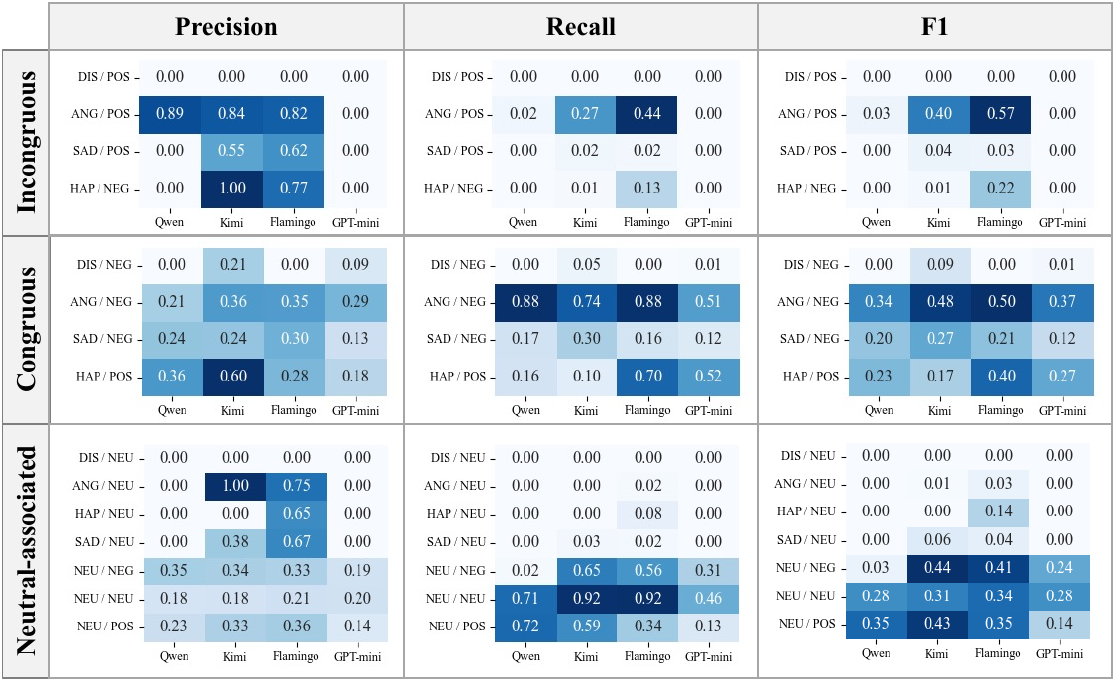}
  \caption {LALM performance across acoustic emotions and semantic sentiment pairs. Each figure label is an abbreviation representing the corresponding acoustic emotion category and semantic sentiment. For example, DIS/POS indicates acoustic disgust and semantic positive.}
    \label{fig:alm_analysis}
\end{figure*}

\subsection{LALM Performance Across Acoustic Emotions and Semantic Sentiments}

We examined dual-condition performance of LALMs, including Qwen2-Audio, Kimi-Audio, Audio-Flamingo3, and GPT-audio-mini across acoustic emotion category and semantic sentiment settings (Figure~\ref{fig:alm_analysis}). These results demonstrate that incongruous cases pose a significant challenge for existing LALMs, which rarely produce predictions in which acoustic and semantic are incongruous, resulting in very low recall and, consequently, very low F1 scores. Incongruous predictions are made only when both acoustic and semantic cues are strong, resulting in relatively high precision. Among the four incongruous scenarios, angry–positive performs best, consistent with previous listening results showing the highest accuracy for angry acoustic cues. Compared to incongruous cases, higher recall observed in congruous cases reflects that LALMs are more likely to produce predictions with consistent semantic and acoustic cues, generally resulting in higher F1 scores across most conditions. As judged by human listeners, disgust is the most difficult emotion to identify, probably because its emotion cues are subtle. For neutral-associated conditions, LALMs struggle particularly with the semantic-neutral setting, where the acoustic signal carries emotion but the semantic content is neutral. In most such cases, the models ignore the acoustic cues and predict neutral–neutral, suggesting a strong bias toward semantic information. Further evidence comes from the acoustic-neutral setting, where models generally achieve higher performance. Overall, results indicate that both acoustic–semantic incongruent cases and semantic-neutral cases are particularly challenging for LALMs.

\subsection{Impact of SFT on LALM Behavior}
\subsubsection{Layer-wise Acoustic-semantic Probing}\label{sec:results_linear_probing}

\begin{figure*}[!h]
    \centering
    % --- Row 1: Base Setting ---
    \begin{subfigure}{0.30\textwidth}
        \includegraphics[width=\linewidth]{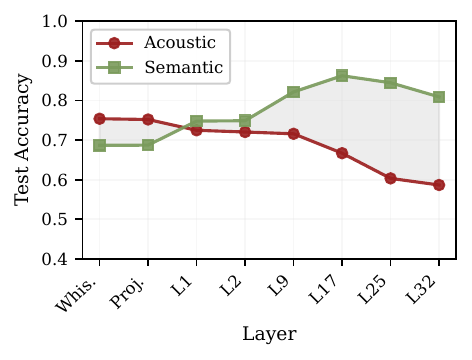}
        \caption{Qwen2-Audio (base)}
        \label{fig:qwen_base}
    \end{subfigure}
    \hfill
    \begin{subfigure}{0.30\textwidth}
        \includegraphics[width=\linewidth]{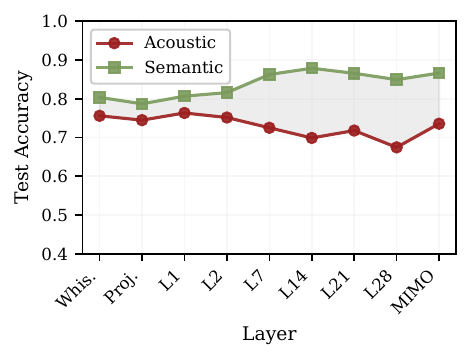}
        \caption{Kimi-Audio (base)}
        \label{fig:kimi_base}
    \end{subfigure}
    \hfill
    \begin{subfigure}{0.30\textwidth}
        \includegraphics[width=\linewidth]{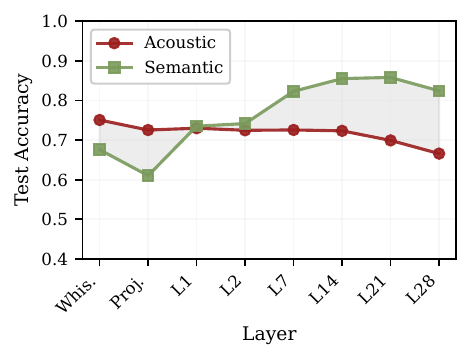}
        \caption{Audio-Flamingo3 (base)}
        \label{fig:af_base}
    \end{subfigure}

    %\vspace{0.5cm} % Vertical spacing between rows

    % --- Row 2: LoRA Setting ---
    \begin{subfigure}{0.30\textwidth}
        \includegraphics[width=\linewidth]{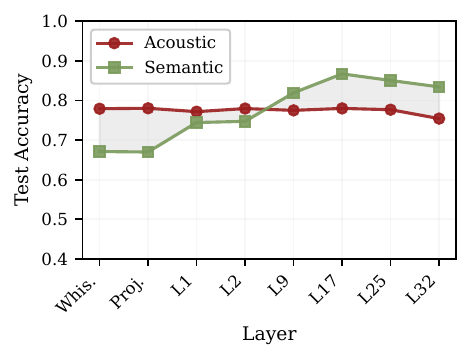}
        \caption{Qwen2-Audio-SFT}
        \label{fig:qwen_lora}
    \end{subfigure}
    \hfill
    \begin{subfigure}{0.30\textwidth}
        \includegraphics[width=\linewidth]{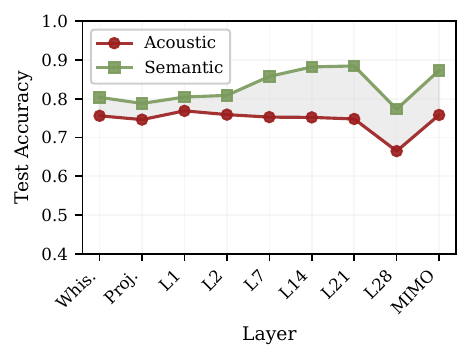}
        \caption{Kimi-Audio-SFT}
        \label{fig:kimi_lora}
    \end{subfigure}
    \hfill
    \begin{subfigure}{0.30\textwidth}
        \includegraphics[width=\linewidth]{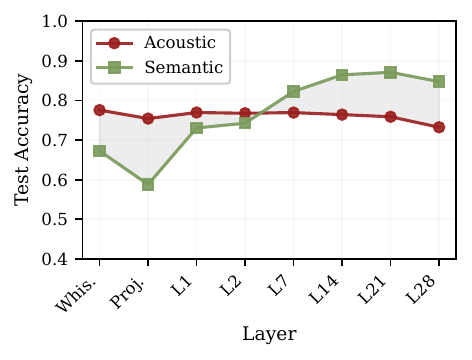}
        \caption{Audio-Flamingo3-SFT}
        \label{fig:af_lora}
    \end{subfigure}
    %\caption{Comparison of acoustic emotion versus semantic sentiment test accuracy across model layers for three models %(Qwen2-Audio Audio-Flamingo3, and Kimi-Audio) 
    %under Base (top) and LoRA (bottom) settings. Shaded areas represent the acoustic-semantic gap.}
    \caption{Layer-wise probing accuracy for acoustic emotion (red) and semantic sentiment (green) across model depth using mean pooling over hidden states. Top row: base models showing acoustic-semantic divergence in deeper layers. Bottom row: SFT models showing reduced divergence with improved acoustic accuracy.
    }
    \label{fig:layer_wise_accuracy_graph}
\end{figure*}

We performed layer-wise linear probing of LALMs before and after SFT. Across all base models (Figure~\ref{fig:layer_wise_accuracy_graph}, top row), acoustic and semantic probing accuracy diverge with depth: semantic accuracy (green) rises above $85\%$ in deeper layers, while acoustic accuracy (red) drops from approximately $75\%$ at the Whisper encoder to approximately $67\%$ (Kimi-Audio, Audio-Flamingo3) and $59\%$ (Qwen2-Audio) at the final LLM layer, suggesting a progressive representational shift from acoustic cues to more semantic-oriented features.
After SFT (Table~\ref{tab:acoustic_semantic_gap_probing} and Figure~\ref{fig:layer_wise_accuracy_graph}, bottom row), acoustic performance improves most where base models were weakest, with gains concentrated in later layers. Improvements range from $5.3$ percentage points (pp) (Kimi-Audio L14) to $17.4$ pp (Qwen2-Audio L25), while early-layer gains remain comparatively modest, resulting in more uniform acoustic accuracy throughout layers.
%acoustic performance primarily improves in deeper layers where base models perform worst, with gains ranging from $5.3$ percentage points (pp) (Kimi-Audio L14) to $17.4$ pp (Qwen2-Audio L25). 
%Improvements increase with depth, with Qwen2-Audio showing modest early gains (L1: $+4.7$ pp) but significant improvements in deep layers (L25: $+17.4$ pp, L32: $+16.8$ pp).

Semantic accuracy after SFT remains mostly similar with changes within $\pm2.5$ pp across nearly all layers and models, with a notable exception in Kimi-Audio L28 ($-7.6$ pp). As this layer precedes Kimi-Audio's MIMO module, we hypothesize that the additional module introduces a distribution shift %This layer precedes Kimi-Audio's MIMO module, and we hypothesize that SFT may redistribute representational emphasis toward the MIMO module, which shows $+2.3$ pp acoustic improvement 
(details in Appendix~\ref{sec:kimi_l28_analysis}). 
This appears unique to Kimi-Audio. Since acoustic gains consistently exceed semantic changes (excluding Kimi-Audio L28), the reduced gap $\Delta G$ in Table~\ref{tab:acoustic_semantic_gap_probing} appears driven primarily by improved acoustic processing rather than semantic degradation.

\begin{table}[h!]
\centering
\small
%\resizebox{\columnwidth}{!}{%
\begin{tabular}{|cccc|}
\hline
 \multicolumn{1}{|c|}{Layer   }  & \multicolumn{1}{c|}{$\Delta Acc_{\text{acou}} \uparrow$ }  & \multicolumn{1}{c|}{ $\Delta Acc_{\text{sem}} \uparrow$}  & \multicolumn{1}{c|}{$\Delta G \downarrow$}  \\ \hline 
\multicolumn{4}{|c|}{\cellcolor[HTML]{EFEFEF}Qwen2-Audio}                                                                                 \\ \hline
 %\multicolumn{1}{|c|}{Projector }  & \multicolumn{1}{c|}{0.028}  & \multicolumn{1}{c|}{-0.017}  & 0.046\\ \hline
 \multicolumn{1}{|c|}{L1  }  & \multicolumn{1}{c|}{0.047}  & \multicolumn{1}{c|}{-0.004}  & 0.004\\ \hline
 \multicolumn{1}{|c|}{L17 }  & \multicolumn{1}{c|}{0.113}  & \multicolumn{1}{c|}{0.005}  & -0.108\\ \hline
  \multicolumn{1}{|c|}{L25 }  & \multicolumn{1}{c|}{0.174}  & \multicolumn{1}{c|}{0.006}  & -0.168\\ \hline
 \multicolumn{1}{|c|}{L32 }  & \multicolumn{1}{c|}{0.168}  & \multicolumn{1}{c|}{0.025}  & -0.143\\
\hline \hline
\multicolumn{4}{|c|}{\cellcolor[HTML]{EFEFEF}Kimi-Audio}                                                                                   \\ \hline
 %\multicolumn{1}{|c|}{Projector }  & \multicolumn{1}{c|}{0.001}  & \multicolumn{1}{c|}{0.001}  & 0.000\\ \hline
  \multicolumn{1}{|c|}{L1  }  & \multicolumn{1}{c|}{0.005}  & \multicolumn{1}{c|}{-0.002}  & -0.008\\ \hline
  \multicolumn{1}{|c|}{L14 }  & \multicolumn{1}{c|}{0.053}  & \multicolumn{1}{c|}{0.003}  & -0.050\\ \hline
 \multicolumn{1}{|c|}{L28 }  & \multicolumn{1}{c|}{-0.010}  & \multicolumn{1}{c|}{-0.076}  & -0.067\\ \hline
  \multicolumn{1}{|c|}{MIMO }  & \multicolumn{1}{c|}{0.023}  & \multicolumn{1}{c|}{0.007}  & -0.016\\
\hline \hline
\multicolumn{4}{|c|}{\cellcolor[HTML]{EFEFEF}Audio-Flamingo3}                                                                                             \\ \hline
 %\multicolumn{1}{|c|}{Projector }  & \multicolumn{1}{c|}{0.029}  & \multicolumn{1}{c|}{-0.022}  & 0.051\\ \hline
  \multicolumn{1}{|c|}{L1  }  & \multicolumn{1}{c|}{0.040}  & \multicolumn{1}{c|}{-0.005}  & 0.034\\ \hline
\multicolumn{1}{|c|}{L14 }  & \multicolumn{1}{c|}{0.041}  & \multicolumn{1}{c|}{0.010}  & -0.032\\ \hline
  \multicolumn{1}{|c|}{L28 }  & \multicolumn{1}{c|}{0.067}  & \multicolumn{1}{c|}{0.023}  & -0.043\\ \hline
\end{tabular}%
\caption{Impact of SFT on layer-wise accuracy and the acoustic-semantic gap $G$ across models. $\Delta$ represents changes ($\text{SFT} - \text{base}$). Negative $\Delta G$ values indicate reduced gap.}
  \label{tab:acoustic_semantic_gap_probing}
\end{table}
These findings suggest that, after SFT on CREMA-ASIS, acoustic emotion information becomes more linearly decodable from deeper LLM representations while semantic sentiment remains comparably accessible, leading to a reduced acoustic-semantic gap across layers. %Importantly, probing results reflect the availability of information in intermediate representations rather than the model's internal decision-making process. Our analysis therefore provides diagnostic evidence that SFT alters the representational balance between acoustic and semantic cues. 
Because probing reflects the information available in intermediate representations rather than the model's decision process, our analysis provides diagnostic evidence that SFT alters the representational balance between acoustic and semantic cues, rather than direct evidence about how the model uses these cues.
Complete probing results for mean pooling and last hidden state pooling, as well as analysis of the Kimi-Audio outlier are presented in Appendix~\ref{sec:additional_probing_results}.

\subsubsection{Performance on CREMA-ASIS Test Set}
Table~\ref{tab:dataset_performance} shows the performance of LALMs on CREMA-ASIS test set. Before SFT, Audio-Flamingo3 consistently outperforms others across all evaluated tasks, demonstrating particularly strong semantic understanding. Despite this, all models show difficulty in correctly predicting both semantic and acoustic labels simultaneously, resulting in low $Acc_{\text{dual}}$. The relatively low $Acc_{\text{acou}}$ of GPT-audio-mini may be primarily attributable to frequent audio processing failures. Specifically, the model fails to process 928 out of 5,879 samples, and these cases are treated as incorrect during evaluation, thereby substantially degrading its overall performance. After SFT, all open-source LALMs exhibit significant performance improvements. To verify whether SFT on CREMA-ASIS affects the models' original capabilities, we also evaluate transcription performance. Results show that WER of the transcription task for LALMs remains stable at approximately 0.07, indicating no noticeable degradation in transcription ability.
\begin{table}[htbp!]
\centering
\small
%\resizebox{\columnwidth}{!}{%
\begin{tabular}{|c|c|c|c|}
\hline
                   & $Acc_{\text{acou}}$ & $Acc_{\text{sem}}$ & $Acc_{\text{dual}}$ \\  \hline
GPT-audio-mini     & 0.174        & 0.716        & 0.153   \\ \hline
Qwen2-Audio         & 0.356        & 0.669        & 0.207   \\ \hline
Kimi-Audio         & 0.421        & 0.722        & 0.285   \\ \hline
Audio-Flamingo3     & 0.426        & 0.780        & 0.320   \\ \hline \hline
\rowcolor[HTML]{EFEFEF} 
Qwen2-Audio-SFT     & 0.783        & 0.876        & 0.683   \\ \hline 
\rowcolor[HTML]{EFEFEF} 
Kimi-Audio-SFT     & 0.752        & 0.896        & 0.673   \\ \hline 
\rowcolor[HTML]{EFEFEF} 
Audio-Flamingo3-SFT & 0.751        & 0.902        & 0.678   \\ \hline 
\end{tabular}%
%}
  \caption{Performance on CREMA-ASIS test set.}
  \label{tab:dataset_performance}
\end{table}

\subsubsection{Evaluation on a Joint Setting}

Unlike the prior section, evaluating LALMs using the same task-specific prompt as in SFT, this assesses the model under a joint recognition setting. LALMs are prompted to perform a single task without explicit specification of the modality or cue to attend to, aiming to assess whether SFT affects the models’ general emotion and sentiment recognition and their robustness to prompts that differ from those used during SFT. In each run, the model is prompted with either “Identify the emotion of the given audio.” or “Identify the sentiment of the given audio.”, without imposing explicit acoustic or semantic constraints. Table~\ref{tab:meld_performance} reports results on the MELD dataset, showing that emotion recognition performance improves across all models. However, for Audio-Flamingo3, sentiment performance decreases despite improved emotion recognition, perhaps due to the relatively limited linguistic diversity in CREMA-ASIS constraining a model originally possessing strong semantic understanding.
\begin{table}[htbp!]
\centering
%\resizebox{\columnwidth}{!}{%
\small
\begin{tabular}{|c|c|c|}
\hline
                   & Emotion & Sentiment \\ \hline
Qwen2-Audio         & 0.518  & 0.545    \\ \hline
\rowcolor[HTML]{EFEFEF} 
Qwen2-Audio-SFT     & \textbf{0.529}  & \textbf{0.585}    \\ \hline \hline
Kimi-Audio   &  0.342\footnotemark & 0.476  \\ \hline
\rowcolor[HTML]{EFEFEF} 
Kimi-Audio-SFT     & \textbf{0.536}  & \textbf{0.542}    \\ \hline \hline
Audio-Flamingo3    & 0.517  & \textbf{0.629}      \\ \hline
\rowcolor[HTML]{EFEFEF} 
Audio-Flamingo3-SFT & \textbf{0.532}  & 0.603    \\ \hline
\end{tabular}%
%}
\caption{Performance on the MELD test set (Acc).}
  \label{tab:meld_performance}
\end{table}

\footnotetext{The difference between our Kimi scores and those reported in the original paper is due to prompt variation: the original study outputs category codes (A, B, C), while we output the actual emotion labels (happy, sad, etc.).}

\subsubsection{Evaluation on Out-of-domain Dataset}\label{sec:results_listen_dataset_evaluation}

We evaluate the generalizability of the fine-tuned models on the out-of-domain LISTEN dataset (Table~\ref{tab:listen_performance}). Results indicate that, after SFT on CREMA-ASIS, both Qwen2-Audio and Audio-Flamingo3 achieve improved acoustic task performance. Performance degradation of Kimi-Audio's $Acc_{\text{acou}}$ may be attributed to the original Kimi model having previously been SFT on RAVDESS~\citep{livingstone2018ryerson}, SAVEE~\citep{jackson2014surrey}, and IEMOCAP~\citep{busso2008iemocap} datasets, which also constitute part of the LISTEN evaluation data. These were not included during our subsequent SFT stage, leading to a shift away from the fitted dataset and thus reduced performance on LISTEN. For the semantic task, SFT Qwen2-Audio also improved semantic performance, while Kimi-Audio and Audio-Flamingo3 decreased. %while the decrease in Kimi-Audio and Audio-Flamingo3’s semantic performance may be attributed to the same factors affecting the joint sentiment task. 
However, we note that  $Acc_{\text{sem}}$ in Table~\ref{tab:listen_performance} is reported using the labels provided by the LISTEN dataset, which may not reflect transcript-level sentiment. We provide further analysis of $Acc_{\text{sem}}$ for this dataset in Appendix~\ref{sec:supplement_listen_dataset}.

\begin{table}[htbp!]
\centering
\small
%\resizebox{\columnwidth}{!}{%
\begin{tabular}{|c|c|c|}
\hline
                   & $Acc_{\text{acou}}$    & $Acc_{\text{sem}}$    \\ \hline
Qwen2-Audio         & 0.515& 0.409\\ \hline
\rowcolor[HTML]{EFEFEF} 
Qwen2-Audio-SFT     & \textbf{0.570}& \textbf{0.470}\\ \hline
Kimi-Audio         & \textbf{0.516} & \textbf{0.565}\\ \hline
\rowcolor[HTML]{EFEFEF} 
Kimi-Audio-SFT     & 0.481& 0.330\\ \hline
Audio-Flamingo3     & 0.607          & \textbf{0.508}\\ \hline
\rowcolor[HTML]{EFEFEF} 
Audio-Flamingo3-SFT & \textbf{0.620}& 0.479\\ \hline
\end{tabular}%
%}
  \caption{Evaluation on out-of-domain data.}
 \label{tab:listen_performance}
\end{table}

%\section{Background}

\section{Conclusion}
We introduce CREMA-ASIS, a synthetic dataset for evaluating LALMs’ bias toward acoustic and semantic affective cues in speech. Subjective tests show that TTS models can reproduce the acoustic emotion of reference audio, enabling controlled speech generation. Using this dataset, we reveal that LALMs struggle with acoustic–semantic incongruence and exhibit a tendency to underweight acoustic cues when semantic content is neutral. Layer-wise probing analysis shows that deeper layers exhibit reduced acoustic decodability while semantic decodability improves, widening the gap between modalities. Fine-tuning on our synthetic data mitigates this by improving the availability of acoustic emotion information in deeper layers while largely preserving semantic sentiment representations, reducing modality disparity. We hope this work draws attention to the fact that speech may inherently encode distinct acoustic and semantic information, and that existing LALMs have limited ability to disentangle the two. We further show that TTS-synthesized data may offer a promising direction to mitigate this limitation.

\section*{Limitations}
\paragraph{Methodology limitations:} %The primary goal of this study is to provide a baseline evaluation of LALMs on the acoustic–semantic inconsistent task. Naturally, this focus entails certain limitations that are beyond the scope of the current work. The SFT training parameters used in this study may not be optimal. Although different models may require distinct hyperparameter configurations to achieve their best performance, all LALMs are fine-tuned under the same settings for consistency. Similarly, alternative training or adaptation strategies beyond standard SFT could potentially further improve performance, but developing such strategies is not the focus of this paper. Additionally, our analysis relies on linear probing, which characterizes the linear accessibility of information in intermediate representations but does not reveal how or whether models internally use this information during inference. 
The primary goal of this study is to provide a baseline evaluation of LALMs on the acoustic–semantic inconsistent task, and certain limitations follow from this scope. 
While we fine-tune all LALMs under the same SFT hyperparameter settings for fair comparison, we acknowledge that these settings may not be optimal for each model. Alternative adaptation strategies beyond standard SFT could further improve performance but are not the focus of this work. Additionally, linear probing reveals the accessibility of information in intermediate representations but not whether models use it during inference.

\paragraph{Dataset quality and evaluation limitations:} %We acknowledge limitations of the proposed dataset. First, we do not have perceived emotion labels for our dataset, except for the 150 samples used in the subjective evaluation to verify the TTS model’s ability. The labels in our dataset correspond to intended emotion, as they are derived from reference audio produced by actors who intended to express specific emotions. Intended and perceived emotion may naturally diverge, as listeners may not always accurately perceive others’ emotional states, and perception may be unavoidably influenced by the semantic content. Second, the synthetic data may contain artifacts that are difficult for humans to perceive and could potentially affect learning. This possibility exists even though our manual listening suggests the generated audio sounds natural, and the listening task, conducted without annotators having prior exposure to the audios and the labels, revealed patterns consistent with the original audio references. Lastly, we note that two of the three evaluators in this task are co-authors while the third was not involved in dataset construction. Although agreement between the independent annotator and co-authors was consistent, and all evaluators were blinded to the labels, a fully independent evaluation panel would further strengthen confidence in these results. 
We acknowledge several limitations of the proposed dataset. 
First, except for the 150 samples used in the subjective evaluation to verify the TTS model's ability, the labels correspond to intended rather than perceived emotion, as they are derived from actors who aimed to express specific emotions. Intended and perceived emotion may diverge, as perception can be influenced by both listener subjectivity and semantic content. %Intended and perceived emotion may diverge, and listeners may not always accurately perceive others' emotional states, and perception may  be unavoidably influenced by semantic content. 
Second, although our listening task, conducted with evaluators blinded to labels and audio references, revealed patterns consistent with the original references, our synthetic data may contain imperceptible artifacts that could affect learning.
Lastly, two of the three evaluators are co-authors, with the third not being involved in dataset construction. Agreement between the independent annotator and co-authors was consistent, but a fully independent panel would further strengthen confidence.

\paragraph{Generalizability and scope limitations:} %Our dataset is generated from acted emotion references, which are often exaggerated and easier to identify than spontaneous emotions. Models trained on acted emotions may not generalize well to spontaneous emotion recognition. To evaluate generalizability, we test SFT models on MELD, derived from TV shows, and the LISTEN dataset, which includes acted and spontaneous subsets not used in SFT. These experiments suggest that our method exhibits some generalization ability, but potential limitations remain. Notably, our experiments show that LALMs struggle with acoustic-semantic inconsistencies even for synthesized acted speech. We therefore view the ability to detect emotions in synthesized acted data as a preliminary step toward recognizing emotions in real-world spontaneous speech. We believe our work makes a meaningful contribution by highlighting significant failures of current LALMs in handling acoustic-semantic inconsistencies, providing solid empirical evidence of their limitations, and suggesting directions for mitigation. Although synthetic data can be generated in principle without limit, the performance gains they provide are inherently bounded. During training, we observed that LALMs may start to overfit before completing a full pass through the training set, suggesting that the training set may contain redundant information. Lastly, CREMA-ASIS uses only a subset of sentences from the GoEmotions dataset and focuses on the basic emotion categories. Future work could aim to select representative samples to create a more informative and efficient training dataset, increase sentence diversity, and investigate acoustic–semantic incongruence in other emotion categories or conditions.
Our dataset is generated from acted emotion references, which are often exaggerated and easier to identify than spontaneous speech.
To assess generalizability, we evaluate SFT models on MELD (TV dialogue) and LISTEN (acted and spontaneous subsets), finding some transferability, but potential limitations remain. Notably, our results show that LALMs struggle with acoustic-semantic inconsistencies even for synthesized acted speech, so we view this as a necessary first step towards handling real-world spontaneous emotion. Although synthetic data can be generated without limit, we observe overfitting before a full epoch, suggesting diminishing returns from redundant training samples. Lastly, CREMA-ASIS covers only a subset of GoEmotions sentences and basic emotion categories. Future work could increase sentence diversity and explore additional emotion categories.

%Although synthetic data can be generated without limit, performance gains are bounded as we observe overfitting before a full training epoch, suggesting redundancy in the training set. Lastly, CREMA-ASIS covers only a subset of GoEmotions sentences and basic emotion categories. Future work could increase sentence diversity and explore additional emotion categories.

%\section*{Acknowledgments}
\section*{Acknowledgments}
This work was supported by the National Institutes of Health under Grant No. 1R01AG081928-01. We thank other members of the Columbia School of Nursing for their fruitful discussion, including Pallavi Gupta and Zhihong Zhang, as well as members of VNS Health, including Sasha Vergez and Margaret McDonald. We also thank Amazon Web Services (AWS) for providing computing credits that supported the development of this work.

\bibliography{custom}

\newpage

\appendix

\section{Experiment Parameter Settings}

\subsection{LALM Prompt}
\label{sec:sft_prompt}
\begin{figure}[htpb!]
  \includegraphics[width=\columnwidth]{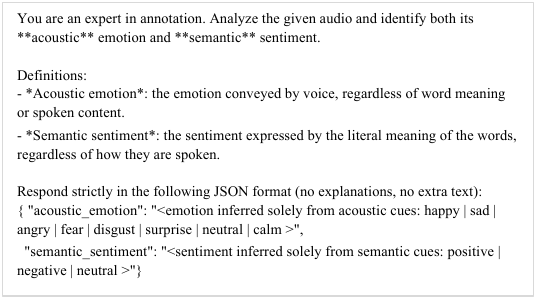}
  \caption{LALM prompt, instructing LALM to simultaneously predict acoustic emotion categories and semantic sentiment labels, with acoustic predictions including categories beyond those in CREMA-ASIS.}
  \label{fig:prompt_sft_instruction}
\end{figure}

Figure~\ref{fig:prompt_sft_instruction} shows the prompt used for SFT and evaluation of the LALMs. Notice that the prompt includes the acoustic emotion category fear while it does not appear in our CREMA-ASIS dataset. We found that the fear recordings in CREMA-D often lacked naturalness and failed to convey the intended acoustic cues of fear. Therefore, we decided to exclude it from the synthetic dataset. However, we still included the fear category in the prompt because datasets, such as MELD, contain it. This ensures that our evaluation of MELD remains consistent with previous studies, and that the raw LALM performance matches that reported in the original LALM papers.

\subsection{Sentence Selection Prompt}
\label{sec:sentence_selection_prompt}
\begin{figure}[htpb!]
  \includegraphics[width=\columnwidth]{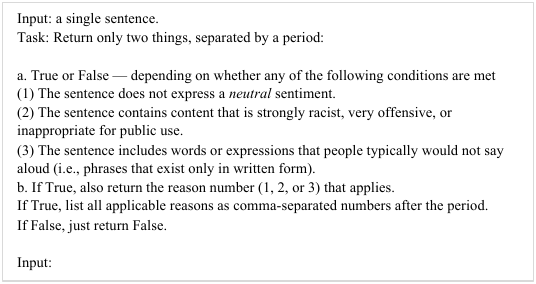}
  \caption{Prompt for sentence selection, where \emph{neutral} is replaced with \emph{positive} or \emph{negative} when filtering sentences with corresponding sentiment.}
  \label{fig:prompt_sentence_Selection}
\end{figure}

Figure~\ref{fig:prompt_sentence_Selection} shows the prompt used for text sentence filtering. 

\subsection{IndexTTS2 Generation Settings}
\label{sec:indextts2_settings}

Synthetic speech is generated using IndexTTS2~\cite{zhou2026indextts2} in a zero-shot, speaker-conditioned setting. For each sample, a reference audio clip from the original CREMA-D corpus is used as the speaker prompt for voice cloning, and the target sentence from the filtered GoEmotions is passed as the text input. The audio clips are preprocessed using SEMamba~\cite{chao2024investigation}~\footnote{\url{https://github.com/RoyChao19477/SEMamba/blob/main/ckpts/vd.pth}} with the \emph{vd.pth} checkpoint for speech enhancement. We control acoustic emotions solely based on the reference voice and set the emotion vector to all zeros and disable the textual emotion prompt. Standard FP32 precision is used, and all outputs are written as WAV files.

\subsection{Acoustic Filtering settings}
\label{sec:acoustic_filtering_settings}

The goal of acoustic filtering with AER is to remove samples whose conveyed emotion is clearly different from the target emotion. However, the AER model is not perfect and may fail to identify the correct emotion. If we retained only samples whose AER predictions exactly matched the target emotion, many valid samples would be removed even though the synthesized audio conveys an emotion similar to the reference audio. For example, sad audio is often identified as neutral by the AER, and it rarely predicts the sad label; even when it does, the audio often does not clearly convey sad emotion. Table~\ref{tab:AER_matching} shows the target–prediction pairs retained in this study.

\begin{table}[htbp!]
\centering
\small
%\resizebox{\columnwidth}{!}{%
\begin{tabular}{|c|c|}
\hline
Audio Reference & AER Prediction            \\ \hline \hline
happy           & happy, surprised, neutral \\ \hline
neutral         & neutral, angry            \\ \hline
sad             & sad, neutral                   \\ \hline
disgust         & surprised, neutral        \\ \hline
angry           & angry, surprised, neutral \\ \hline
\end{tabular}%
%}
  \caption{Target–prediction pairs retained.}
 \label{tab:AER_matching}
\end{table}

\subsection{SFT Parameter Settings}
\label{sec:SFT_setting}

The models are fine-tuned using a rank-stabilized LoRA (rsLoRA) adapter with rank $r=4$ and $\text{LoRA}_\alpha=32$, applied to all linear layers in the \texttt{CAUSAL\_LM} setting. We use a learning rate of $1 \times 10^{-5}$ with a gradient accumulation step of 4 with LoRA dropout of 0.1. Checkpoints are saved every 2,500 steps, and the checkpoint achieving the best overall performance is selected. SFT is performed on NVIDIA L40, RTX A5500, and A100-40GB GPUs.

\subsection{Layer-wise Linear Probing Training Parameters Settings}
\label{sec:experiment_setup_probing}
We use the data split from Table~\ref{tab:dataset_statistic}. We train all linear probes for 20 epochs using Adam~\cite{kingma2014adam} with cosine annealing~\cite{loshchilov2016sgdr}, grid-searching over learning rates $\{5\times10^{-2}, 1\times10^{-2}, 5\times10^{-3}, 1\times10^{-3}\}$ evaluated on the validation set. Each probe consists of a single linear layer and uses input dimensions matching the representation size (1280 for Whisper, 4096 for Qwen2-Audio, 3584 for Kimi-Audio and Audio-Flamingo3) with output dimensions of 5 and 3 for the acoustic emotion and semantic sentiment tasks respectively, which corresponds to the categories. We use no hidden layers in any of our probing experiments. %The random seeds used for our 6 runs are 0, 1, 2, 3, 4 and 42.

\subsection{LISTEN Dataset Setup}\label{sec:listen_dataset_setup_Additional}
%We divide the data samples from LISTEN into two subsets for acoustic emotion analysis and semantic sentiment analysis based on the question associated with each sample. Samples are assigned to the acoustic emotion analysis subset if their associated question asks about emotion conveyed through vocal prosody or tone (e.g., vocal expression, emotional tone of the speaker's voice), and to the semantic sentiment analysis subset if their associated question asks about emotion conveyed through the literal textual content (e.g., semantic meaning of the words used). The full set of questions used for each subset is listed in Table~\ref{tab:question_split_listen}.
%For the semantic analysis, we convert the emotion labels in the LISTEN dataset to positive, negative, and neutral sentiments following the mapping in Table~\ref{tab:sentiment_map_listen}
We split the LISTEN dataset into an acoustic emotion analysis subset and a semantic sentiment analysis subset based on the question associated with each sample from the original dataset. A sample is assigned to the acoustic subset if the question asks about emotion conveyed through vocal prosody or tone (e.g., vocal expression, emotional tone of the speaker's voice), and to the semantic subset if the question asks about emotion conveyed through the literal textual content (e.g., the semantics of the spoken words). The full set of questions used for each subset is listed in Table~\ref{tab:question_split_listen}.

When constructing the semantic subset, we need to be careful about a property of the LISTEN dataset. LISTEN assigns to each utterance a single emotion label taken from its source corpus, and in conditions where acoustic and lexical cues are treated as aligned, this same label is attached to the text-centric question (i.e., the LISTEN question within our semantic subset) as well. Because these source labels (e.g., in IEMOCAP) were annotated from audio or audiovisual delivery, they encode the emotion of the vocal performance rather than the semantics from the transcript alone. We further discuss the consequences of this in Appendix~\ref{sec:supplement_listen_dataset}. 

For the semantic analysis, we map the resulting labels to positive, negative, and neutral sentiment following Table~\ref{tab:sentiment_map_listen}, and exclude samples with ambiguous mappings.

\begin{table*}[htpb!]
\centering
\small
\renewcommand{\arraystretch}{1.2}
\begin{tabular}{|p{0.47\textwidth}|p{0.47\textwidth}|}
\hline
\multicolumn{1}{|c|}{\textbf{Acoustic Emotion Analysis}} & \multicolumn{1}{c|}{\textbf{Semantic Sentiment Analysis}} \\ \hline
What emotion is communicated through the speaker's vocal prosody? &
Based solely on the text content, what emotion would you identify? \\ \hline
Based on the vocal expression, what emotion is the speaker feeling? &
From the semantic content alone, what emotion is being expressed? \\ \hline
How would you describe the emotional tone of the speaker? &
What emotional tone is conveyed by the literal meaning of this statement? \\ \hline
Listening to the voice, what emotion is being expressed? &
What feeling is suggested by the meaning of these words? \\ \hline
What emotion is expressed in the speaker's voice? &
Based on the content of this text, what emotion would the person likely be feeling? \\ \hline
What emotional state is reflected in the speaker's voice? &
Reading this text, what emotional state does the speaker appear to be in? \\ \hline
What emotion does the speaker convey through their tone? &
What emotion is conveyed by the words in this statement? \\ \hline
\end{tabular}%
\caption{Questions used to split LISTEN samples into acoustic emotion analysis and semantic sentiment analysis subsets.}
\label{tab:question_split_listen}
\end{table*}

\begin{table}[htbp!]
\centering
\small
\begin{tabular}{|c|p{5.5cm}|}
\hline
Sentiment & Original Labels \\ \hline \hline
positive  & amusement, excitement, happiness, happy, pleasantly surprised \\ \hline
negative  & anger, angry, annoyance, anxious, concern, concerned, confusion, contempt, disappointment, disgust, fear, frustration, ridicule, sad, sadness \\ \hline
neutral   & calm, neutral \\ \hline
\end{tabular}%
\caption{Semantic sentiment label mapping from the LISTEN dataset to our setting.}
\label{tab:sentiment_map_listen}
\end{table}

\section{Supplementary Analysis of LISTEN Dataset}\label{sec:supplement_listen_dataset}
In Section~\ref{sec:results_listen_dataset_evaluation}, we observed low $Acc_{\text{sem}}$ on LISTEN, and reduced $Acc_{\text{sem}}$ after SFT for Kimi-Audio and Audio-Flamingo3, when scored against the labels provided by LISTEN. Upon manual inspection, this behavior appears to stem primarily from how LISTEN's labels are constructed rather than from the models themselves. LISTEN assigns each utterance a single emotion label taken from its source corpus, and attaches this label to the text-centric (``semantic'') question as well. For corpora such as IEMOCAP, these labels were annotated from audio or audiovisual delivery, so for utterances whose transcript's sentiment is neutral or divergent from the vocal delivery, the text-centric label reflects the acoustic emotion rather than the sentiment expressed by the transcript. This makes these labels unsuitable targets for a purely text-based sentiment task.

Several examples within the ``Matched-Emotion'' for text setting (LISTEN's experiment type 2A) illustrate this mismatch. The transcript ``Somebody please call 911'' paired with the question ``What emotion is conveyed by the words in this statement?'' is labeled \emph{happiness}, whereas the words alone convey distress or concern rather than the positive sentiment the label implies. Furthermore, ``I would've loved to'' and ``As a matter of fact, that's perfectly true'' are labeled \emph{sadness} and \emph{anger} respectively, while in both cases the text alone reads more neutral-to-positive. Rather than annotation errors, these examples are labeled based on the delivered acoustic emotions while being associated with a text-centric question under the assumption that lexical and acoustic emotional cues match. However, for these examples, the assumption does not hold, and the text-centric label and the transcript's semantic sentiment diverge.

To quantify this divergence, we feed the transcripts of the audios alone to a strong text-only classifier, Qwen3-32B~\citep{yang2025qwen3}, and compare its semantic sentiment predictions against LISTEN's text-centric labels using the mapping in Table~\ref{tab:sentiment_map_listen}. We treat Qwen3-32B's predictions here as a proxy-label for transcript sentiment, not as ground truth.
On the three transcript examples above, Qwen3-32B predicts negative, positive, and positive, consistent with what a reader would infer from the transcript alone. However, across the full semantic subset (651 samples), Qwen3-32B's predictions agree with LISTEN's labels on only 36.9\% of samples. This low agreement suggests that the text-centric labels encode acoustic affect rather than lexical. When we instead score the three LALMs' semantic predictions from Table~\ref{tab:listen_performance} (which were inferred \emph{from the audio}) against Qwen3-32B's \emph{text-only} transcript-level predictions (see Table~\ref{tab:listen_semantic_performance_qwen3_32b_pseudolabel}), their agreement with transcript-level sentiment is substantially higher than against LISTEN's labels.
This effect is most pronounced for Kimi-Audio, whose agreement increases after SFT on CREMA-ASIS (0.594 to 0.731), with smaller increases for Qwen2-Audio (0.622 to 0.662) and Audio-Flamingo3 (0.711 to 0.720). Notably, Kimi-Audio's semantic score moves in opposite directions depending only on the scoring reference. It drops under LISTEN's labels but rises under transcript-level references, even though the model's predictions are identical.

While our results suggest that SFT improves the models' alignment with transcript-level sentiment rather than degrading it, establishing a true semantic sentiment gold-label would require manual human annotation of the dataset, which we leave to future work. Therefore, we treat the weaker LISTEN $Acc_{\text{sem}}$ in Table~\ref{tab:listen_performance} with caution.

\begin{table}[h!]
\centering
\small
%\resizebox{\columnwidth}{!}{%
\begin{tabular}{|c|c|}
\hline
                   & Agreement with Qwen3-32B     \\ \hline \hline
Qwen2-Audio         & 0.622\\ \hline
\rowcolor[HTML]{EFEFEF} 
Qwen2-Audio-SFT     & \textbf{0.662}\\ \hline
Kimi-Audio         & 0.594\\ \hline
\rowcolor[HTML]{EFEFEF} 
Kimi-Audio-SFT     & \textbf{0.731}\\ \hline
Audio-Flamingo3     & 0.711\\ \hline
\rowcolor[HTML]{EFEFEF} 
Audio-Flamingo3-SFT & \textbf{0.720}\\ \hline
\end{tabular}%
%}
  \caption{Agreement between each LALM's semantic predictions from audio inputs and Qwen3-32B's transcript-level sentiment predictions on LISTEN, before and after SFT on our CREMA-ASIS. Note that this is a reference-agreement measure, not accuracy against verified ground truth. }
 \label{tab:listen_semantic_performance_qwen3_32b_pseudolabel}
\end{table}

%With the label mapping from Table~\ref{tab:sentiment_map_listen} as targets, we perform simple text-only sentiment classification of the transcripts of the LISTEN dataset using Qwen3-32B~\citep{yang2025qwen3}. For the three examples above, the model predicted negative, positive, and positive, respectively, matching the common consensus. However, the overall accuracy across all samples was merely 36.9\%. Surprisingly, as shown in Table~\ref{tab:listen_semantic_performance_qwen3_32b_pseudolabel}, when using Qwen3-32B's predictions as target labels, all three LALMs studied yield significantly higher $Acc_{\text{sem}}$ for the LISTEN dataset. Moreover, each model's $Acc_{\text{sem}}$ also improves after SFT on CREMA-ASIS. While a complete human inspection of the entire LISTEN dataset is required to confirm the labeling quality, the LALMs' semantic sentiment alignment with Qwen3-32B shows signs further signs 

%“He had a car accident.”

\section{Supplementary Layer-Wise Probing Results}
\label{sec:additional_probing_results}

\subsection{Complete Probing Results}
Table~\ref{tab:complete_acoustic_semantic_table} shows the complete results of our layer-wise probing experiments across three models. In addition to the changes between base and SFT models (columns denoted with $\Delta$), we also report the acoustic emotion ($Acc_{\text{acou}}^{\text{base}}, Acc_{\text{acou}}^{\text{SFT}}$), semantic sentiment accuracies ($Acc_{\text{sem}}^{\text{base}}, Acc_{\text{sem}}^{\text{SFT}}$), and acoustic-semantic gaps ($G^{\text{base}}, G^{\text{SFT}}$) for each layer probed across base and SFT models. 

\begin{figure*}[!h]
    \centering
    % --- Row 1: Base Setting ---
    \begin{subfigure}{0.30\textwidth}
        \includegraphics[width=\linewidth]{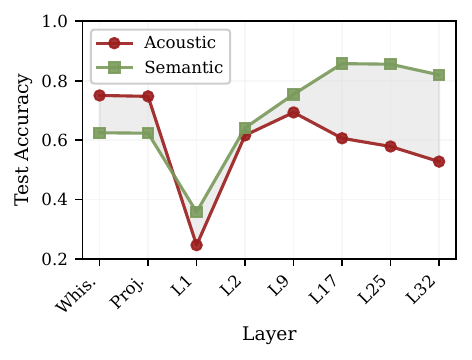}
        \caption{Qwen2-Audio (base)}
        \label{fig:qwen_base}
    \end{subfigure}
    \hfill
    \begin{subfigure}{0.30\textwidth}
        \includegraphics[width=\linewidth]{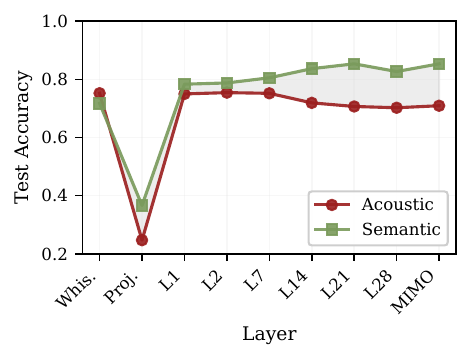}
        \caption{Kimi-Audio (base)}
        \label{fig:kimi_base}
    \end{subfigure}
    \hfill
    \begin{subfigure}{0.30\textwidth}
        \includegraphics[width=\linewidth]{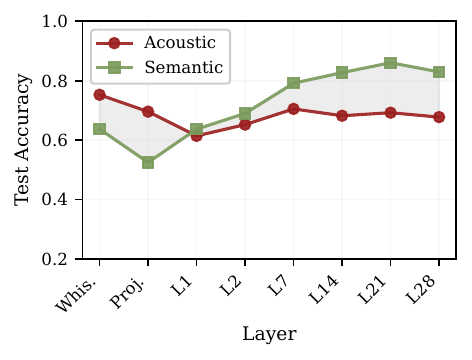}
        \caption{Audio-Flamingo3 (base)}
        \label{fig:af_base}
    \end{subfigure}

    %\vspace{0.5cm} % Vertical spacing between rows

    % --- Row 2: LoRA Setting ---
    \begin{subfigure}{0.30\textwidth}
        \includegraphics[width=\linewidth]{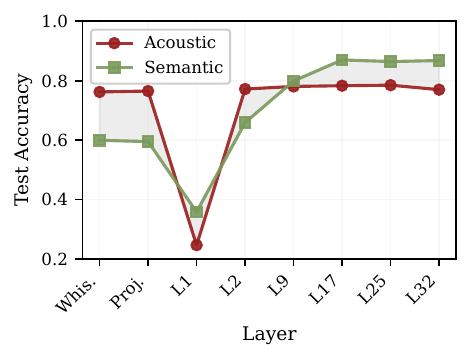}
        \caption{Qwen2-Audio-SFT}
        \label{fig:qwen_lora}
    \end{subfigure}
    \hfill
    \begin{subfigure}{0.30\textwidth}
        \includegraphics[width=\linewidth]{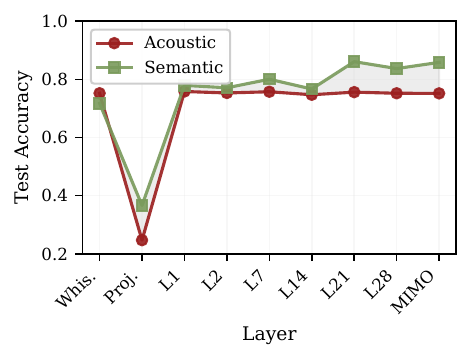}
        \caption{Kimi-Audio-SFT}
        \label{fig:kimi_lora}
    \end{subfigure}
    \hfill
    \begin{subfigure}{0.30\textwidth}
        \includegraphics[width=\linewidth]{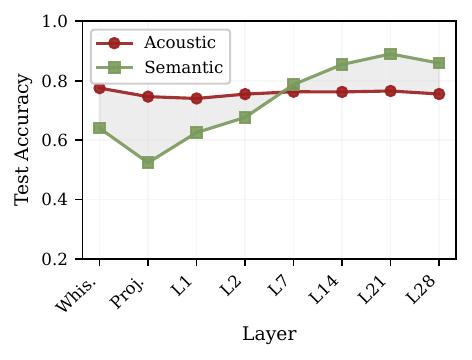}
        \caption{Audio-Flamingo3-SFT}
        \label{fig:af_lora}
    \end{subfigure}
    %\caption{Comparison of acoustic emotion versus semantic sentiment test accuracy across model layers for three models %(Qwen2-Audio Audio-Flamingo3, and Kimi-Audio) 
    %under Base (top) and LoRA (bottom) settings. Shaded areas represent the acoustic-semantic gap.}
    \caption{Layer-wise probing accuracy for acoustic emotion (red) and semantic sentiment (green) across model depth using the last hidden states. Top row: base models showing acoustic-semantic divergence in deeper layers. Bottom row: SFT models showing reduced divergence with improved acoustic accuracy.
    }
    \label{fig:layer_wise_accuracy_graph_last_pooling}
\end{figure*}

Table~\ref{tab:complete_acoustic_semantic_table} also presents the same experiment using the last hidden state as features in place of mean pooled hidden states while Figure~\ref{fig:layer_wise_accuracy_graph_last_pooling} shows the layer-wise trajectories for both acoustic emotion and semantic sentiment accuracies. We observe similar behavior for the last hidden state pooling as mean pooling: for base models, the acoustic-semantic gap widens the deeper we propagate through the layers as acoustic accuracies decrease while semantic accuracies increase. Similar to before, we observe that SFT improves the acoustic emotion accuracy the most where the base models performed the worst while the changes are modest in earlier layers, leading to more uniform acoustic performance throughout the layers. Again, semantic accuracy changes comparatively less with SFT, leading to gap reductions being largely driven by acoustic gains.

\begin{table*}
\centering
\small
%\resizebox{\columnwidth}{!}{%
\begin{tabular}{|ccccccccccc|}
\hline
\multicolumn{1}{|c|}{Layer   }  & \multicolumn{1}{c|}{Pooling} & \multicolumn{1}{c|}{ $Acc_{\text{acou}}^{\text{base}}$ }   & \multicolumn{1}{c|}{ $Acc_{\text{sem}}^{\text{base}}$ }  & \multicolumn{1}{c|}{$ G^{\text{base}} $} & \multicolumn{1}{|c|}{$Acc_{\text{acou}}^{\text{SFT}}$    }  & \multicolumn{1}{|c|}{$Acc_{\text{sem}}^{\text{SFT}}$    } & \multicolumn{1}{|c|}{  $G^{\text{SFT}}$   }  & \multicolumn{1}{|c|}{$\Delta Acc_{\text{acou}} \uparrow$ } & \multicolumn{1}{|c|}{ $\Delta Acc_{\text{sem}} \uparrow$} & \multicolumn{1}{|c|}{$\Delta G \downarrow$}     \\ \hline \hline
\multicolumn{11}{|c|}{\cellcolor[HTML]{EFEFEF}Qwen2-Audio}                                                                                 \\ \hline
\multicolumn{1}{|c|}{Whisper   }  &   \multicolumn{1}{|c|}{Mean}  & \multicolumn{1}{|c|}{0.754}   & \multicolumn{1}{|c|}{0.687}   & \multicolumn{1}{|c|}{0.067}  & \multicolumn{1}{|c|}{0.779}   & \multicolumn{1}{|c|}{0.671}   &\multicolumn{1}{|c|}{0.108}   & \multicolumn{1}{c|}{0.025}  & \multicolumn{1}{c|}{-0.016}  & 0.041\\ \hline
\multicolumn{1}{|c|}{Projector }  &   \multicolumn{1}{|c|}{Mean}  & \multicolumn{1}{|c|}{0.752}   & \multicolumn{1}{|c|}{0.687}   & \multicolumn{1}{|c|}{0.065}  & \multicolumn{1}{|c|}{0.780}   & \multicolumn{1}{|c|}{0.670}   &\multicolumn{1}{|c|}{0.111}   & \multicolumn{1}{c|}{0.028}  & \multicolumn{1}{c|}{-0.017}  & 0.046\\ \hline
\multicolumn{1}{|c|}{L1}          &   \multicolumn{1}{|c|}{Mean}          & \multicolumn{1}{|c|}{0.725}         & \multicolumn{1}{|c|}{0.748}         & \multicolumn{1}{|c|}{0.023}         & \multicolumn{1}{|c|}{0.772}         & \multicolumn{1}{|c|}{0.744}         &\multicolumn{1}{|c|}{0.027}         & \multicolumn{1}{c|}{0.047}  & \multicolumn{1}{c|}{-0.004}  & 0.004\\ \hline
\multicolumn{1}{|c|}{L2}          &   \multicolumn{1}{|c|}{Mean}          & \multicolumn{1}{|c|}{0.721}         & \multicolumn{1}{|c|}{0.749}         & \multicolumn{1}{|c|}{0.029}   & \multicolumn{1}{|c|}{0.780}         & \multicolumn{1}{|c|}{0.747}         &\multicolumn{1}{|c|}{0.032}         & \multicolumn{1}{c|}{0.059}  & \multicolumn{1}{c|}{-0.002}  & 0.004\\ \hline
\multicolumn{1}{|c|}{L9}          &   \multicolumn{1}{|c|}{Mean}          & \multicolumn{1}{|c|}{0.716}         & \multicolumn{1}{|c|}{0.822}         & \multicolumn{1}{|c|}{0.106}   & \multicolumn{1}{|c|}{0.775}         & \multicolumn{1}{|c|}{0.820}         &\multicolumn{1}{|c|}{0.044}         & \multicolumn{1}{c|}{0.059}  & \multicolumn{1}{c|}{-0.002}  & -0.062\\ \hline
\multicolumn{1}{|c|}{L17}         &   \multicolumn{1}{|c|}{Mean}         & \multicolumn{1}{|c|}{0.667}         & \multicolumn{1}{|c|}{0.863}         & \multicolumn{1}{|c|}{0.196}   & \multicolumn{1}{|c|}{0.780}         & \multicolumn{1}{|c|}{0.867}         &\multicolumn{1}{|c|}{0.087}         & \multicolumn{1}{c|}{0.113}  & \multicolumn{1}{c|}{0.005}  & -0.108\\ \hline
\multicolumn{1}{|c|}{L25}         &   \multicolumn{1}{|c|}{Mean}         & \multicolumn{1}{|c|}{0.603}         & \multicolumn{1}{|c|}{0.845}         & \multicolumn{1}{|c|}{0.242}   & \multicolumn{1}{|c|}{0.777}         & \multicolumn{1}{|c|}{0.851}         &\multicolumn{1}{|c|}{0.074}         & \multicolumn{1}{c|}{0.174}  & \multicolumn{1}{c|}{0.006}  & -0.168\\ \hline
\multicolumn{1}{|c|}{L32}         &   \multicolumn{1}{|c|}{Mean}         & \multicolumn{1}{|c|}{0.587}         & \multicolumn{1}{|c|}{0.809}         & \multicolumn{1}{|c|}{0.223}   & \multicolumn{1}{|c|}{0.754}         & \multicolumn{1}{|c|}{0.835}         &\multicolumn{1}{|c|}{0.080}         & \multicolumn{1}{c|}{0.168}  & \multicolumn{1}{c|}{0.025}  & -0.143\\
\hline \hline
\multicolumn{1}{|c|}{Whisper   }  &   \multicolumn{1}{|c|}{Last}  & \multicolumn{1}{|c|}{0.750}   & \multicolumn{1}{|c|}{0.625}   & \multicolumn{1}{|c|}{0.126}  & \multicolumn{1}{|c|}{0.762}   & \multicolumn{1}{|c|}{0.599}   &\multicolumn{1}{|c|}{0.163}   & \multicolumn{1}{c|}{0.012}  & \multicolumn{1}{c|}{-0.025}  & 0.037\\ \hline
\multicolumn{1}{|c|}{Projector }  &   \multicolumn{1}{|c|}{Last}  & \multicolumn{1}{|c|}{0.747}   & \multicolumn{1}{|c|}{0.623}   & \multicolumn{1}{|c|}{0.124}  & \multicolumn{1}{|c|}{0.765}   & \multicolumn{1}{|c|}{0.595}   &\multicolumn{1}{|c|}{0.170}   & \multicolumn{1}{c|}{0.018}  & \multicolumn{1}{c|}{-0.028}  & 0.046\\ \hline
\multicolumn{1}{|c|}{L1}          &   \multicolumn{1}{|c|}{Last}          & \multicolumn{1}{|c|}{0.246}         & \multicolumn{1}{|c|}{0.358}         & \multicolumn{1}{|c|}{0.112}         & \multicolumn{1}{|c|}{0.246}         & \multicolumn{1}{|c|}{0.358}         &\multicolumn{1}{|c|}{0.112}         & \multicolumn{1}{c|}{0.000}  & \multicolumn{1}{c|}{0.000}  & 0.000\\ \hline
\multicolumn{1}{|c|}{L2}          &   \multicolumn{1}{|c|}{Last}          & \multicolumn{1}{|c|}{0.616}         & \multicolumn{1}{|c|}{0.641}         & \multicolumn{1}{|c|}{0.025}   & \multicolumn{1}{|c|}{0.772}         & \multicolumn{1}{|c|}{0.659}         &\multicolumn{1}{|c|}{0.113}         & \multicolumn{1}{c|}{0.156}  & \multicolumn{1}{c|}{0.018}  & 0.088\\ \hline
\multicolumn{1}{|c|}{L9}          &   \multicolumn{1}{|c|}{Last}          & \multicolumn{1}{|c|}{0.693}         & \multicolumn{1}{|c|}{0.754}         & \multicolumn{1}{|c|}{0.061}   & \multicolumn{1}{|c|}{0.781}         & \multicolumn{1}{|c|}{0.799}         &\multicolumn{1}{|c|}{0.019}         & \multicolumn{1}{c|}{0.087}  & \multicolumn{1}{c|}{0.045}  & -0.043\\ \hline
\multicolumn{1}{|c|}{L17}         &   \multicolumn{1}{|c|}{Last}         & \multicolumn{1}{|c|}{0.606}         & \multicolumn{1}{|c|}{0.858}         & \multicolumn{1}{|c|}{0.251}   & \multicolumn{1}{|c|}{0.783}         & \multicolumn{1}{|c|}{0.870}         &\multicolumn{1}{|c|}{0.087}         & \multicolumn{1}{c|}{0.177}  & \multicolumn{1}{c|}{0.012}  & -0.164\\ \hline
\multicolumn{1}{|c|}{L25}         &   \multicolumn{1}{|c|}{Last}         & \multicolumn{1}{|c|}{0.579}         & \multicolumn{1}{|c|}{0.856}         & \multicolumn{1}{|c|}{0.277}   & \multicolumn{1}{|c|}{0.785}         & \multicolumn{1}{|c|}{0.865}         &\multicolumn{1}{|c|}{0.080}         & \multicolumn{1}{c|}{0.207}  & \multicolumn{1}{c|}{0.009}  & -0.198\\ \hline
\multicolumn{1}{|c|}{L32}         &   \multicolumn{1}{|c|}{Last}         & \multicolumn{1}{|c|}{0.528}         & \multicolumn{1}{|c|}{0.820}         & \multicolumn{1}{|c|}{0.292}   & \multicolumn{1}{|c|}{0.770}         & \multicolumn{1}{|c|}{0.868}         &\multicolumn{1}{|c|}{0.098}         & \multicolumn{1}{c|}{0.242}  & \multicolumn{1}{c|}{0.048}  & -0.194\\
\hline \hline
\multicolumn{11}{|c|}{\cellcolor[HTML]{EFEFEF}Kimi-Audio}                                                                                   \\ \hline
\multicolumn{1}{|c|}{Whisper   } & \multicolumn{1}{|c|}{Mean} &  \multicolumn{1}{|c|}{0.756}  &  \multicolumn{1}{|c|}{0.804}  &  \multicolumn{1}{|c|}{0.048}  &  \multicolumn{1}{|c|}{0.756}  &  \multicolumn{1}{|c|}{0.804}  & \multicolumn{1}{|c|}{0.048}  & \multicolumn{1}{c|}{0.000}  & \multicolumn{1}{c|}{0.000}  & 0.000\\ \hline
\multicolumn{1}{|c|}{Projector } & \multicolumn{1}{|c|}{Mean} &  \multicolumn{1}{|c|}{0.745}  &  \multicolumn{1}{|c|}{0.787}  &  \multicolumn{1}{|c|}{0.042}  &  \multicolumn{1}{|c|}{0.746}  &  \multicolumn{1}{|c|}{0.788}  & \multicolumn{1}{|c|}{0.042}  & \multicolumn{1}{c|}{0.001}  & \multicolumn{1}{c|}{0.001}  & 0.000\\ \hline
\multicolumn{1}{|c|}{L1}         & \multicolumn{1}{|c|}{Mean}         &  \multicolumn{1}{|c|}{0.763}        &  \multicolumn{1}{|c|}{0.807}        &  \multicolumn{1}{|c|}{0.043} &  \multicolumn{1}{|c|}{0.769}        &  \multicolumn{1}{|c|}{0.805}        & \multicolumn{1}{|c|}{0.036}        & \multicolumn{1}{c|}{0.005}  & \multicolumn{1}{c|}{-0.002}  & -0.008\\ \hline
\multicolumn{1}{|c|}{L2}         & \multicolumn{1}{|c|}{Mean}         &  \multicolumn{1}{|c|}{0.752}        &  \multicolumn{1}{|c|}{0.816}        &  \multicolumn{1}{|c|}{0.064} &  \multicolumn{1}{|c|}{0.759}        &  \multicolumn{1}{|c|}{0.809}        & \multicolumn{1}{|c|}{0.050}        & \multicolumn{1}{c|}{0.007}  & \multicolumn{1}{c|}{-0.007}  & -0.015\\ \hline
\multicolumn{1}{|c|}{L7}         & \multicolumn{1}{|c|}{Mean}         &  \multicolumn{1}{|c|}{0.725}        &  \multicolumn{1}{|c|}{0.863}        &  \multicolumn{1}{|c|}{0.138} &  \multicolumn{1}{|c|}{0.753}        &  \multicolumn{1}{|c|}{0.858}        & \multicolumn{1}{|c|}{0.105}        & \multicolumn{1}{c|}{0.028}  & \multicolumn{1}{c|}{-0.005}  & -0.033\\ \hline
\multicolumn{1}{|c|}{L14}        & \multicolumn{1}{|c|}{Mean}        &  \multicolumn{1}{|c|}{0.699}        &  \multicolumn{1}{|c|}{0.879}        &  \multicolumn{1}{|c|}{0.180} &  \multicolumn{1}{|c|}{0.752}        &  \multicolumn{1}{|c|}{0.882}        & \multicolumn{1}{|c|}{0.130}        & \multicolumn{1}{c|}{0.053}  & \multicolumn{1}{c|}{0.003}  & -0.050\\ \hline
\multicolumn{1}{|c|}{L21}        & \multicolumn{1}{|c|}{Mean}        &  \multicolumn{1}{|c|}{0.718}        &  \multicolumn{1}{|c|}{0.866}        &  \multicolumn{1}{|c|}{0.148} &  \multicolumn{1}{|c|}{0.748}        &  \multicolumn{1}{|c|}{0.885}        & \multicolumn{1}{|c|}{0.137}        & \multicolumn{1}{c|}{0.030}  & \multicolumn{1}{c|}{0.019}  & -0.011\\ \hline
\multicolumn{1}{|c|}{L28}        & \multicolumn{1}{|c|}{Mean}        &  \multicolumn{1}{|c|}{0.675}        &  \multicolumn{1}{|c|}{0.850}        &  \multicolumn{1}{|c|}{0.175} &  \multicolumn{1}{|c|}{0.665}        &  \multicolumn{1}{|c|}{0.773}        & \multicolumn{1}{|c|}{0.108}        & \multicolumn{1}{c|}{-0.010}  & \multicolumn{1}{c|}{-0.076}  & -0.067\\ \hline
\multicolumn{1}{|c|}{MIMO }      & \multicolumn{1}{|c|}{Mean}      &  \multicolumn{1}{|c|}{0.736}       &  \multicolumn{1}{|c|}{0.867}       &  \multicolumn{1}{|c|}{0.131}   &  \multicolumn{1}{|c|}{0.758}       &  \multicolumn{1}{|c|}{0.873}       & \multicolumn{1}{|c|}{0.115}       & \multicolumn{1}{c|}{0.023}  & \multicolumn{1}{c|}{0.007}  & -0.016\\
\hline \hline
\multicolumn{1}{|c|}{Whisper   } &  \multicolumn{1}{|c|}{Last}  &  \multicolumn{1}{|c|}{0.753}  &  \multicolumn{1}{|c|}{0.718}  &  \multicolumn{1}{|c|}{0.035}  &  \multicolumn{1}{|c|}{0.753}  &  \multicolumn{1}{|c|}{0.718}  & \multicolumn{1}{|c|}{0.035}  & \multicolumn{1}{c|}{0.000}  & \multicolumn{1}{c|}{0.000}  & 0.000\\ \hline
\multicolumn{1}{|c|}{Projector } &  \multicolumn{1}{|c|}{Last}  &  \multicolumn{1}{|c|}{0.247}  &  \multicolumn{1}{|c|}{0.366}  &  \multicolumn{1}{|c|}{0.120}  &  \multicolumn{1}{|c|}{0.247}  &  \multicolumn{1}{|c|}{0.366}  & \multicolumn{1}{|c|}{0.120}  & \multicolumn{1}{c|}{0.000}  & \multicolumn{1}{c|}{0.000}  & 0.000\\ \hline
\multicolumn{1}{|c|}{L1}         &  \multicolumn{1}{|c|}{Last}        &  \multicolumn{1}{|c|}{0.750}        &  \multicolumn{1}{|c|}{0.784}        &  \multicolumn{1}{|c|}{0.034} &  \multicolumn{1}{|c|}{0.759}        &  \multicolumn{1}{|c|}{0.780}        & \multicolumn{1}{|c|}{0.021}        & \multicolumn{1}{c|}{0.008}  & \multicolumn{1}{c|}{-0.004}  & -0.012\\ \hline
\multicolumn{1}{|c|}{L2}         &  \multicolumn{1}{|c|}{Last}        &  \multicolumn{1}{|c|}{0.755}        &  \multicolumn{1}{|c|}{0.788}        &  \multicolumn{1}{|c|}{0.033} &  \multicolumn{1}{|c|}{0.753}        &  \multicolumn{1}{|c|}{0.771}        & \multicolumn{1}{|c|}{0.018}        & \multicolumn{1}{c|}{-0.001}  & \multicolumn{1}{c|}{-0.017}  & -0.015\\ \hline
\multicolumn{1}{|c|}{L7}         &  \multicolumn{1}{|c|}{Last}        &  \multicolumn{1}{|c|}{0.752}        &  \multicolumn{1}{|c|}{0.806}        &  \multicolumn{1}{|c|}{0.053} &  \multicolumn{1}{|c|}{0.758}        &  \multicolumn{1}{|c|}{0.801}        & \multicolumn{1}{|c|}{0.043}        & \multicolumn{1}{c|}{0.005}  & \multicolumn{1}{c|}{-0.005}  & -0.010\\ \hline
\multicolumn{1}{|c|}{L14}        &  \multicolumn{1}{|c|}{Last}        &  \multicolumn{1}{|c|}{0.719}        &  \multicolumn{1}{|c|}{0.837}        &  \multicolumn{1}{|c|}{0.118} &  \multicolumn{1}{|c|}{0.747}        &  \multicolumn{1}{|c|}{0.767}        & \multicolumn{1}{|c|}{0.020}        & \multicolumn{1}{c|}{0.028}  & \multicolumn{1}{c|}{-0.070}  & -0.098\\ \hline
\multicolumn{1}{|c|}{L21}        &  \multicolumn{1}{|c|}{Last}        &  \multicolumn{1}{|c|}{0.707}        &  \multicolumn{1}{|c|}{0.854}        &  \multicolumn{1}{|c|}{0.147} &  \multicolumn{1}{|c|}{0.756}        &  \multicolumn{1}{|c|}{0.861}        & \multicolumn{1}{|c|}{0.105}        & \multicolumn{1}{c|}{0.049}  & \multicolumn{1}{c|}{0.007}  & -0.042\\ \hline
\multicolumn{1}{|c|}{L28}        &  \multicolumn{1}{|c|}{Last}        &  \multicolumn{1}{|c|}{0.703}        &  \multicolumn{1}{|c|}{0.827}        &  \multicolumn{1}{|c|}{0.125} &  \multicolumn{1}{|c|}{0.752}        &  \multicolumn{1}{|c|}{0.837}        & \multicolumn{1}{|c|}{0.085}        & \multicolumn{1}{c|}{0.050}  & \multicolumn{1}{c|}{0.010}  & -0.040\\ \hline
\multicolumn{1}{|c|}{MIMO }      &  \multicolumn{1}{|c|}{Last}       &  \multicolumn{1}{|c|}{0.710}       &  \multicolumn{1}{|c|}{0.854}       &  \multicolumn{1}{|c|}{0.144}   &  \multicolumn{1}{|c|}{0.752}       &  \multicolumn{1}{|c|}{0.859}       & \multicolumn{1}{|c|}{0.107}       & \multicolumn{1}{c|}{0.043}  & \multicolumn{1}{c|}{0.005}  & -0.038\\
\hline \hline
\multicolumn{11}{|c|}{\cellcolor[HTML]{EFEFEF}Audio-Flamingo3}                                                                                             \\ \hline
\multicolumn{1}{|c|}{Whisper   }  &   \multicolumn{1}{|c|}{Mean}  & \multicolumn{1}{|c|}{0.751}   & \multicolumn{1}{|c|}{0.676}   & \multicolumn{1}{|c|}{0.075} & \multicolumn{1}{|c|}{0.776}   & \multicolumn{1}{|c|}{0.673}   &\multicolumn{1}{|c|}{0.103}  & \multicolumn{1}{c|}{0.025}  & \multicolumn{1}{c|}{-0.003}  & 0.028\\ \hline
\multicolumn{1}{|c|}{Projector }  &   \multicolumn{1}{|c|}{Mean}  & \multicolumn{1}{|c|}{0.726}   & \multicolumn{1}{|c|}{0.610}   & \multicolumn{1}{|c|}{0.115} & \multicolumn{1}{|c|}{0.754}   & \multicolumn{1}{|c|}{0.588}   &\multicolumn{1}{|c|}{0.166}  & \multicolumn{1}{c|}{0.029}  & \multicolumn{1}{c|}{-0.022}  & 0.051\\ \hline
\multicolumn{1}{|c|}{L1}          &   \multicolumn{1}{|c|}{Mean}          & \multicolumn{1}{|c|}{0.730}         & \multicolumn{1}{|c|}{0.735}         & \multicolumn{1}{|c|}{0.005}         & \multicolumn{1}{|c|}{0.770}         & \multicolumn{1}{|c|}{0.731}         &\multicolumn{1}{|c|}{0.039}  & \multicolumn{1}{c|}{0.040}  & \multicolumn{1}{c|}{-0.005}  & 0.034\\ \hline
\multicolumn{1}{|c|}{L2}          &   \multicolumn{1}{|c|}{Mean}          & \multicolumn{1}{|c|}{0.725}         & \multicolumn{1}{|c|}{0.742}         & \multicolumn{1}{|c|}{0.017}         & \multicolumn{1}{|c|}{0.768}         & \multicolumn{1}{|c|}{0.743}         &\multicolumn{1}{|c|}{0.025}  & \multicolumn{1}{c|}{0.043}  & \multicolumn{1}{c|}{0.001}  & 0.008\\ \hline
\multicolumn{1}{|c|}{L7}          &   \multicolumn{1}{|c|}{Mean}          & \multicolumn{1}{|c|}{0.726}         & \multicolumn{1}{|c|}{0.823}         & \multicolumn{1}{|c|}{0.098}         & \multicolumn{1}{|c|}{0.770}         & \multicolumn{1}{|c|}{0.823}         &\multicolumn{1}{|c|}{0.053}  & \multicolumn{1}{c|}{0.044}  & \multicolumn{1}{c|}{0.000}  & -0.044\\ \hline
\multicolumn{1}{|c|}{L14}         &   \multicolumn{1}{|c|}{Mean}         & \multicolumn{1}{|c|}{0.723}         & \multicolumn{1}{|c|}{0.855}         & \multicolumn{1}{|c|}{0.132}         & \multicolumn{1}{|c|}{0.764}         & \multicolumn{1}{|c|}{0.865}         &\multicolumn{1}{|c|}{0.101}  & \multicolumn{1}{c|}{0.041}  & \multicolumn{1}{c|}{0.010}  & -0.032\\ \hline
\multicolumn{1}{|c|}{L21}         &   \multicolumn{1}{|c|}{Mean}         & \multicolumn{1}{|c|}{0.699}         & \multicolumn{1}{|c|}{0.858}         & \multicolumn{1}{|c|}{0.159}         &  \multicolumn{1}{|c|}{0.759}         & \multicolumn{1}{|c|}{0.871}         &\multicolumn{1}{|c|}{0.112}  & \multicolumn{1}{c|}{0.060}  & \multicolumn{1}{c|}{0.013}  & -0.047\\ \hline
\multicolumn{1}{|c|}{L28}         &   \multicolumn{1}{|c|}{Mean}         & \multicolumn{1}{|c|}{0.666}         & \multicolumn{1}{|c|}{0.825}         & \multicolumn{1}{|c|}{0.159}         & \multicolumn{1}{|c|}{0.732}         & \multicolumn{1}{|c|}{0.848}         &\multicolumn{1}{|c|}{0.115}  & \multicolumn{1}{c|}{0.067}  & \multicolumn{1}{c|}{0.023}  & -0.043\\ \hline \hline
\multicolumn{1}{|c|}{Whisper   }  &   \multicolumn{1}{|c|}{Last}  & \multicolumn{1}{|c|}{0.753}   & \multicolumn{1}{|c|}{0.637}   & \multicolumn{1}{|c|}{0.116} & \multicolumn{1}{|c|}{0.775}   & \multicolumn{1}{|c|}{0.640}   &\multicolumn{1}{|c|}{0.135}  & \multicolumn{1}{c|}{0.022}  & \multicolumn{1}{c|}{0.003}  & 0.020\\ \hline
\multicolumn{1}{|c|}{Projector }  &   \multicolumn{1}{|c|}{Last}  & \multicolumn{1}{|c|}{0.696}   & \multicolumn{1}{|c|}{0.525}   & \multicolumn{1}{|c|}{0.171} & \multicolumn{1}{|c|}{0.746}   & \multicolumn{1}{|c|}{0.524}   &\multicolumn{1}{|c|}{0.223}  & \multicolumn{1}{c|}{0.051}  & \multicolumn{1}{c|}{-0.001}  & 0.051\\ \hline
\multicolumn{1}{|c|}{L1}          &   \multicolumn{1}{|c|}{Last}          & \multicolumn{1}{|c|}{0.614}         & \multicolumn{1}{|c|}{0.637}         & \multicolumn{1}{|c|}{0.023}         & \multicolumn{1}{|c|}{0.740}         & \multicolumn{1}{|c|}{0.625}         &\multicolumn{1}{|c|}{0.115}  & \multicolumn{1}{c|}{0.126}  & \multicolumn{1}{c|}{-0.011}  & 0.092\\ \hline
\multicolumn{1}{|c|}{L2}          &   \multicolumn{1}{|c|}{Last}          & \multicolumn{1}{|c|}{0.652}         & \multicolumn{1}{|c|}{0.690}         & \multicolumn{1}{|c|}{0.038}         & \multicolumn{1}{|c|}{0.755}         & \multicolumn{1}{|c|}{0.677}         &\multicolumn{1}{|c|}{0.078}  & \multicolumn{1}{c|}{0.103}  & \multicolumn{1}{c|}{-0.013}  & 0.041\\ \hline
\multicolumn{1}{|c|}{L7}          &   \multicolumn{1}{|c|}{Last}          & \multicolumn{1}{|c|}{0.705}         & \multicolumn{1}{|c|}{0.792}         & \multicolumn{1}{|c|}{0.087}         & \multicolumn{1}{|c|}{0.763}         & \multicolumn{1}{|c|}{0.787}         &\multicolumn{1}{|c|}{0.024}  & \multicolumn{1}{c|}{0.058}  & \multicolumn{1}{c|}{-0.004}  & -0.062\\ \hline
\multicolumn{1}{|c|}{L14}         &   \multicolumn{1}{|c|}{Last}         & \multicolumn{1}{|c|}{0.682}         & \multicolumn{1}{|c|}{0.827}         & \multicolumn{1}{|c|}{0.145}         & \multicolumn{1}{|c|}{0.762}         & \multicolumn{1}{|c|}{0.855}         &\multicolumn{1}{|c|}{0.093}  & \multicolumn{1}{c|}{0.080}  & \multicolumn{1}{c|}{0.027}  & -0.053\\ \hline
\multicolumn{1}{|c|}{L21}         &   \multicolumn{1}{|c|}{Last}         & \multicolumn{1}{|c|}{0.692}         & \multicolumn{1}{|c|}{0.861}         & \multicolumn{1}{|c|}{0.168}         &  \multicolumn{1}{|c|}{0.765}         & \multicolumn{1}{|c|}{0.890}         &\multicolumn{1}{|c|}{0.125}  & \multicolumn{1}{c|}{0.073}  & \multicolumn{1}{c|}{0.030}  & -0.043\\ \hline
\multicolumn{1}{|c|}{L28}         &   \multicolumn{1}{|c|}{Last}         & \multicolumn{1}{|c|}{0.678}         & \multicolumn{1}{|c|}{0.830}         & \multicolumn{1}{|c|}{0.152}         & \multicolumn{1}{|c|}{0.755}         & \multicolumn{1}{|c|}{0.859}         &\multicolumn{1}{|c|}{0.104}  & \multicolumn{1}{c|}{0.078}  & \multicolumn{1}{c|}{0.030}  & -0.048\\ \hline
\end{tabular}%
\caption{Layer-wise acoustic emotion accuracy $Acc_{\text{acou}}$, semantic sentiment accuracy $Acc_{\text{sem}}$, and acoustic-semantic gap $G$  between base and SFT LALMs from linear probing experiments. $\Delta$ represents changes ($\text{SFT} - \text{base}$). Negative $\Delta G$ values indicate reduced gap.}
  \label{tab:complete_acoustic_semantic_table}
\end{table*}

\subsection{Kimi-Audio L28 Analysis}
\label{sec:kimi_l28_analysis}
\begin{figure*}[!h]
    \centering
    \begin{subfigure}{0.4\textwidth}
        \includegraphics[width=\linewidth]{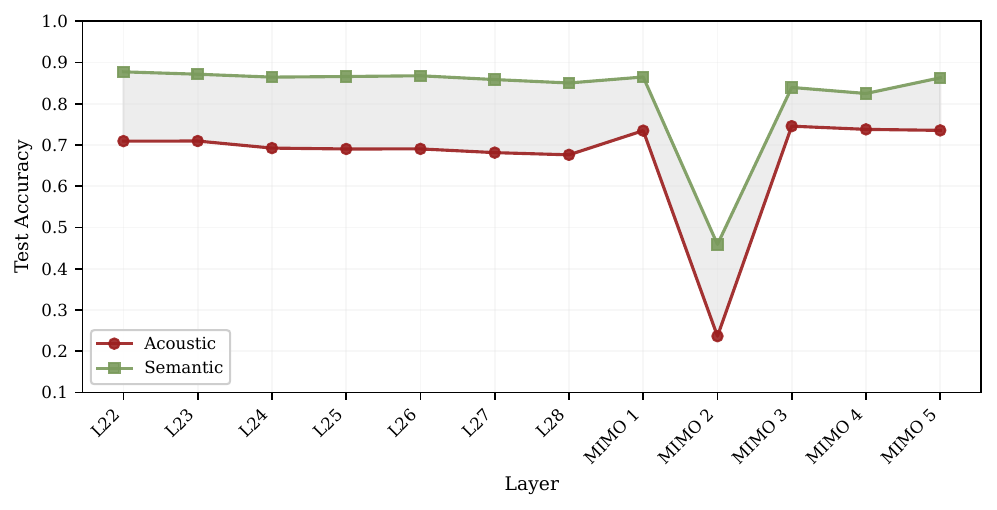}
        \caption{Kimi-Audio (base)}
        \label{fig:kimi_base}
    \end{subfigure}
    \hspace{1cm}
    %\vspace{0.5cm} % Vertical spacing between rows
    \begin{subfigure}{0.4\textwidth}
        \includegraphics[width=\linewidth]{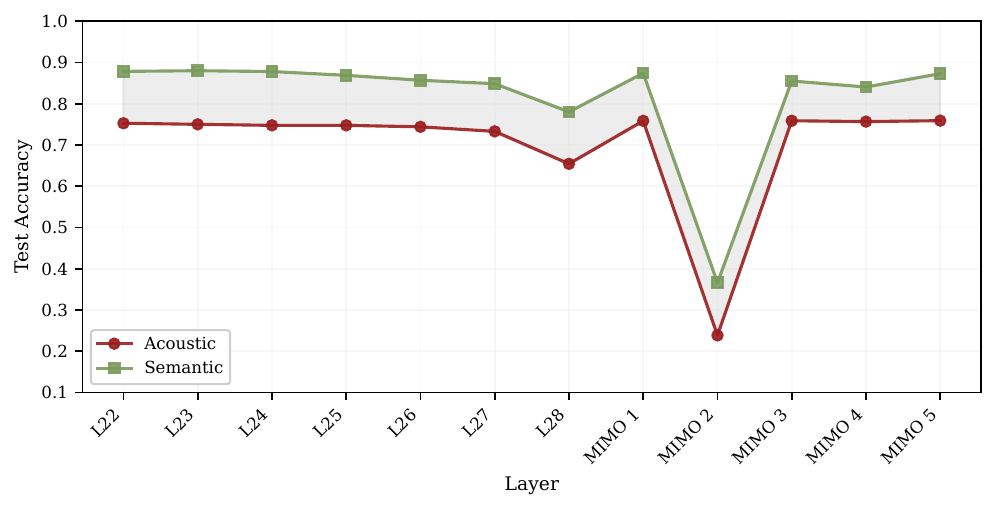}
        \caption{Kimi-Audio (Fine-tuned)}
        \label{fig:kimi_lora}
    \end{subfigure}
    %\caption{Comparison of acoustic emotion versus semantic sentiment test accuracy across model layers for three models %(Qwen2-Audio Audio-Flamingo3, and Kimi-Audio) 
    %under Base (top) and LoRA (bottom) settings. Shaded areas represent the acoustic-semantic gap.}
    \caption{Layer-wise probing accuracy for acoustic emotion (red) and semantic sentiment (green) across model depth using the mean pooling for Kimi-Audio's late LLM layers and MIMO module. Left: base models. Right: SFT models. Accuracies are computed based on means from five runs with different random seeds.
    }
    \label{fig:layer_wise_probing_kimi_audio}
\end{figure*}

To further investigate the accuracy drop of Kimi-Audio L28 from Section~\ref{sec:results_linear_probing}, we extend our probing experiment to layers 22--28 of Kimi-Audio's LLM backbone and layers 1--5 of the MIMO module. We run all experiments with 5 different random seeds and present the mean accuracies in Table~\ref{tab:kimi_audio_probe_table} and Figure~\ref{fig:layer_wise_probing_kimi_audio}. We omit reporting standard deviations as they are all $<0.005$. For the base model, we see a steady decrease in acoustic accuracy through each layer of the LLM backbone followed by a sharp increase at the MIMO module, reaching levels comparable to early LLM layers. The same pattern can be observed for the SFT model, but this time, the drop at Kimi-Audio L28 is significantly sharper. Since the LLM backbone was initially pre-trained on text-only tasks, we hypothesize that there is a distribution mismatch between it and the MIMO module (consisting of a 1B audio transformers), which leads SFT to exacerbate the mismatch. We leave further investigation for future work. 
%Similar to observations from Section~\ref{sec:results_linear_probing}, we see that the acoustic 

Similar to Kimi-Audio L28, we also observe dips in accuracy with the last hidden state pooling for Qwen2-Audio L1, Kimi-Audio Projector (Table~\ref{tab:complete_acoustic_semantic_table} and Figure~\ref{fig:layer_wise_accuracy_graph_last_pooling}) and Kimi-Audio MIMO 2 (Table~\ref{tab:kimi_audio_probe_table} and Figure~\ref{fig:layer_wise_probing_kimi_audio}) at a much larger magnitude. For Qwen2-Audio L1 and Kimi-Audio Projector, SFT has no effect on the accuracy, while for Kimi-Audio L28 and MIMO 2 SFT either reduce the accuracy or barely increase it by 0.2 percentage points. While not conclusive, we note that all drops happen at the very start or end of the LLM layers, which may indicate domain mismatches between the modules. We leave further investigation for future work.

%Kimi-Audio L28 with mean pooling shows anomalous behavior with $-1.0$ pp acoustic and $-7.6$ pp semantic degradation after SFT, contrasting with the consistent improvements observed in other layers and models. We attribute this to Kimi-Audio's unique architecture as L28 is the final transformer layer in the backbone LLM before the additional MIMO module. Looking at Table~\ref{tab:kimi_audio_probe_table} and Figure\ref{fig:layer_wise_probing_kimi_audio}, we observe that MIMO layers

%The MIMO module performs final modality integration and shows $+2.3$ pp acoustic and +$0.7$ pp semantic gains. (3) Fine-tuning may shift representation optimization from L28 to MIMO, specializing L28 for MIMO input rather than direct classification.

%Importantly, this does not invalidate our findings: (1) the MIMO module itself shows acoustic improvement without semantic loss, and (2) this pattern is unique to Kimi-Audio's specialized architecture and absent in standard transformers (Qwen2-Audio, Audio-Flamingo3).

\begin{table*}
\centering
\small
%\resizebox{\columnwidth}{!}{%
\begin{tabular}{|cccccccccc|}
\hline
\multicolumn{10}{|c|}{\cellcolor[HTML]{EFEFEF}Kimi-Audio}                                                                                 \\ \hline
\multicolumn{1}{|c|}{Layer   }  & \multicolumn{1}{c|}{$Acc_{\text{acou}}^{\text{base}}$ }  & \multicolumn{1}{c|}{ $Acc_{\text{sem}}^{\text{base}}$ }  & \multicolumn{1}{c|}{$ G^{\text{base}} $} & \multicolumn{1}{|c|}{$Acc_{\text{acou}}^{\text{SFT}}$    }  & \multicolumn{1}{|c|}{$Acc_{\text{sem}}^{\text{SFT}}$    } & \multicolumn{1}{|c|}{  $G^{\text{SFT}}$   }  & \multicolumn{1}{|c|}{$\Delta Acc_{\text{acou}} \uparrow$ } & \multicolumn{1}{|c|}{ $\Delta Acc_{\text{sem}} \uparrow$} & \multicolumn{1}{|c|}{$\Delta G \downarrow$}   \\ \hline
\multicolumn{1}{|c|}{L22}   & \multicolumn{1}{|c|}{0.710}   & \multicolumn{1}{|c|}{0.878}   & \multicolumn{1}{|c|}{0.168}  & \multicolumn{1}{|c|}{0.753}   & \multicolumn{1}{|c|}{0.879}   &\multicolumn{1}{|c|}{0.126}   & \multicolumn{1}{c|}{0.043}  & \multicolumn{1}{c|}{0.001}  & -0.042\\ \hline
\multicolumn{1}{|c|}{L23}   & \multicolumn{1}{|c|}{0.710}   & \multicolumn{1}{|c|}{0.872}   & \multicolumn{1}{|c|}{0.162}  & \multicolumn{1}{|c|}{0.750}   & \multicolumn{1}{|c|}{0.880}   &\multicolumn{1}{|c|}{0.130}   & \multicolumn{1}{c|}{0.041}  & \multicolumn{1}{c|}{0.009}  & -0.032\\ \hline
\multicolumn{1}{|c|}{L24}         & \multicolumn{1}{|c|}{0.693}         & \multicolumn{1}{|c|}{0.865}         & \multicolumn{1}{|c|}{0.172}         & \multicolumn{1}{|c|}{0.748}         & \multicolumn{1}{|c|}{0.878}         &\multicolumn{1}{|c|}{0.131}         & \multicolumn{1}{c|}{0.055}  & \multicolumn{1}{c|}{0.014}  & -0.042\\ \hline
\multicolumn{1}{|c|}{L25}         & \multicolumn{1}{|c|}{0.691}         & \multicolumn{1}{|c|}{0.866}         & \multicolumn{1}{|c|}{0.176}   & \multicolumn{1}{|c|}{0.748}         & \multicolumn{1}{|c|}{0.869}         &\multicolumn{1}{|c|}{0.121}         & \multicolumn{1}{c|}{0.057}  & \multicolumn{1}{c|}{0.003}  & -0.054\\ \hline
\multicolumn{1}{|c|}{L26}         & \multicolumn{1}{|c|}{0.691}         & \multicolumn{1}{|c|}{0.868}         & \multicolumn{1}{|c|}{0.177}   & \multicolumn{1}{|c|}{0.744}         & \multicolumn{1}{|c|}{0.857}         &\multicolumn{1}{|c|}{0.113}         & \multicolumn{1}{c|}{0.054}  & \multicolumn{1}{c|}{-0.011}  & -0.065\\ \hline
\multicolumn{1}{|c|}{L27}         & \multicolumn{1}{|c|}{0.682}         & \multicolumn{1}{|c|}{0.859}         & \multicolumn{1}{|c|}{0.177}   & \multicolumn{1}{|c|}{0.733}         & \multicolumn{1}{|c|}{0.849}         &\multicolumn{1}{|c|}{0.116}         & \multicolumn{1}{c|}{0.052}  & \multicolumn{1}{c|}{-0.010}  & -0.062\\ \hline
\multicolumn{1}{|c|}{L28}         & \multicolumn{1}{|c|}{0.676}         & \multicolumn{1}{|c|}{0.851}         & \multicolumn{1}{|c|}{0.174}   & \multicolumn{1}{|c|}{0.654}         & \multicolumn{1}{|c|}{0.780}         &\multicolumn{1}{|c|}{0.126}         & \multicolumn{1}{c|}{-0.022}  & \multicolumn{1}{c|}{-0.070}  & -0.048\\ \hline
\multicolumn{1}{|c|}{MIMO 1}         & \multicolumn{1}{|c|}{0.735}         & \multicolumn{1}{|c|}{0.865}         & \multicolumn{1}{|c|}{0.130}   & \multicolumn{1}{|c|}{0.759}         & \multicolumn{1}{|c|}{0.874}         &\multicolumn{1}{|c|}{0.116}         & \multicolumn{1}{c|}{0.024}  & \multicolumn{1}{c|}{0.009}  & -0.014\\
\hline
\multicolumn{1}{|c|}{MIMO 2}  &  \multicolumn{1}{|c|}{0.236}  &  \multicolumn{1}{|c|}{0.459}  &  \multicolumn{1}{|c|}{0.223}  &  \multicolumn{1}{|c|}{0.238}  &  \multicolumn{1}{|c|}{0.366}  & \multicolumn{1}{|c|}{0.128}  & \multicolumn{1}{c|}{0.002}  & \multicolumn{1}{c|}{-0.093}  & -0.095\\ \hline
\multicolumn{1}{|c|}{MIMO 3}  &  \multicolumn{1}{|c|}{0.746}  &  \multicolumn{1}{|c|}{0.840}  &  \multicolumn{1}{|c|}{0.094}  &  \multicolumn{1}{|c|}{0.759}  &  \multicolumn{1}{|c|}{0.855}  & \multicolumn{1}{|c|}{0.097}  & \multicolumn{1}{c|}{0.013}  & \multicolumn{1}{c|}{0.016}  & 0.003\\ \hline
\multicolumn{1}{|c|}{MIMO 4}        &  \multicolumn{1}{|c|}{0.738}        &  \multicolumn{1}{|c|}{0.825}        &  \multicolumn{1}{|c|}{0.087} &  \multicolumn{1}{|c|}{0.757}        &  \multicolumn{1}{|c|}{0.841}        & \multicolumn{1}{|c|}{0.084}        & \multicolumn{1}{c|}{0.019}  & \multicolumn{1}{c|}{0.016}  & -0.003\\ \hline
\multicolumn{1}{|c|}{MIMO 5}        &  \multicolumn{1}{|c|}{0.736}        &  \multicolumn{1}{|c|}{0.863}        &  \multicolumn{1}{|c|}{0.128} &  \multicolumn{1}{|c|}{0.759}        &  \multicolumn{1}{|c|}{0.873}        & \multicolumn{1}{|c|}{0.114}        & \multicolumn{1}{c|}{0.024}  & \multicolumn{1}{c|}{0.010}  & -0.014\\ \hline
\end{tabular}%
\caption{Layer-wise acoustic emotion accuracy $Acc_{\text{acou}}$, semantic sentiment accuracy $Acc_{\text{sem}}$, and acoustic-semantic gap $G$  between base and SFT LALMs from linear probing experiments. $\Delta$ represents changes ($\text{SFT} - \text{base}$). Negative $\Delta G$ values indicate reduced gap. Numbers are computed based on means from five runs with different random seeds.}
  \label{tab:kimi_audio_probe_table}
\end{table*}

\end{document}